\documentclass{article}
\usepackage[hypertexnames=false]{hyperref}
\usepackage{amsmath}
\usepackage{amssymb}
\usepackage{amsthm}
\usepackage{placeins}
\usepackage{wrapfig}
\usepackage{subfigure}
\usepackage{nicefrac}
\usepackage{xcolor}
\usepackage{algorithm,algorithmic}
\usepackage{hyperref}

\newcommand{\E}[2]{\operatorname{\mathbb{E}}_{#1}\left[#2\right]}
\newcommand{\EE}{\mathbb{E}}
\newcommand{\enstq}[2]{\left(#1~\middle|\middle|~#2\right)} 
\usepackage{microtype}
\usepackage{graphicx}
\usepackage{subfigure}
\usepackage{booktabs} %
\usepackage{graphicx, caption, lipsum}
\usepackage{fancyhdr}
\usepackage{movie15}
\usepackage{titlesec}
\titleformat{\section}{\large\bfseries}{\thesection}{0.3em}{}

\usepackage{atbegshi}
\AtBeginDocument{\AtBeginShipoutNext{\AtBeginShipoutDiscard}}

\usepackage[utf8]{inputenc}
\renewenvironment{abstract}
{
	\centerline{\large\bf Abstract}
	\vspace{-0.12in}\begin{quote}}
	{\par\end{quote}\vskip 0.12in}

\begin{document}
	\twocolumn[
	
	\title{\rule{\linewidth}{1.5pt}\\ \textbf{{\LARGE Solving The Dynamic Volatility Fitting Problem:} \\ A Deep Reinforcement Learning Approach
			  } \rule{\linewidth}{1.5pt}}

	\author{Emmanuel Gnabeyeu
		\thanks{emmanuel.gnabeyeu-mbiada@polytechnique.edu} \\ \small London, UK.
		\and
		Omar Karkar \thanks{omar.karkar@citi.com} \\ \small London, UK.  
		\and
		Imad Idboufous \thanks{imad.idboufous@citi.com} \\ \small London, UK.
	}
	
	]

	\maketitle
	\thispagestyle{plain}
	\newpage 
	\pagenumbering{arabic}

	
	\begin{abstract}
		The volatility fitting  is one of the core problems in the equity derivatives business. Through a set of deterministic rules, the degrees of freedom in the implied volatility surface encoding (parametrization, density, diffusion) are defined. Whilst very effective, this approach widespread in the industry is not natively tailored to learn from shifts in market regimes and discover
unsuspected optimal behaviors. In this paper, we change the classical paradigm and apply the latest advances in Deep Reinforcement Learning(DRL) to solve the fitting problem. In particular, we show that variants of Deep Deterministic Policy Gradient (DDPG) and Soft Actor Critic (SAC) can achieve at least as-good as standard fitting algorithms. Furthermore, we explain why the reinforcement learning framework is appropriate to handle complex objective functions and is natively adapted for online learning.

	\end{abstract}
	

	\section{INTRODUCTION}
	
	The implied volatility, which is one of the most important risk factors in the equity business, is usually encoded in a functional form with a limited set of  quasi-orthogonal parameters.
	This functional form, often referred to as volatility parametrization, allows users to describe the dynamic properties (movements and deformations) of the implied volatility surface in an easy way and helps 
	in generating predictive signals for pricing and risk management of equity derivatives.\\
	
	The process through which the coefficients of the volatility parametrization are determined is called calibration and is often performed automatically \footnote{It can also be performed on a less regular basis depending on the liquidity of the asset.}
	by an algorithm tailored to optimize pre-determined objective functions. For instance, most algorithms aim at matching the mid implied volatility at all expiries where there is a visible market under non-arbitrage constraints. 
	The fitting algorithms  used in the industry often account for large sets of conditions such as liquidity, presence of outlier quotes or even earnings and macro events however all those algorithms have native rigidity in them as they follow preset rules 
	for every market conditions. By design, the "deterministic algorithms'' cannot find optimal solutions outside of their definition scope and cannot re-use past accumulated experiences as every output is flashed out.\\
	
	In this paper, we look at the fitting problem from the angle of Reinforcement Learning (RL). In this new paradigm, we assimilate the fitting-algorithms to agents evolving in a stochastic environment where: (a) states are the collection of observable market quotes and 
	anterior volatility surfaces  \footnote{States can also account for spot and/or forward movements.} (b) actions are the bumps to apply in order to shift the prior volatility parametrization and (c) rewards are the opposite of well designed errors. By adopting a RL framework, one can benefit from the native exploration/exploitation trade-off where optimal actions are taken with some margin for exploration that could unveil new optimal decisions. Furthermore, in continuous states (which is the case for the volatility fitting problem), one can heavily benefit from replay buffers (see. \ref{appendix:subsection:ReplayMem} ) which are massive logs with historical or synthetic experiences. Finally, by learning in a stochastic environment where market quotes are continuously evolving RL agents can learn to track market dynamics and better stabilize the volatility parametrization coefficients.\\
	
	Our findings show that the RL framework offers a powerful and adaptative approach to solving the continuous fitting problem. Leveraging Actor-Critic RL based techniques, we observed that the RL agent was capable of adjusting to the evolving market dynamics with minimal intervention, thus offering a more flexible and adaptable solution than traditional deterministic models.  \\
	
	The remainder of the paper is organized as follows. In section \ref{section:vol_fitting_problem} , we present the fitting problem, discuss traditional approaches and introduce our main RL alternative. Section \ref{section:basics_rl} provides some foundational concepts and background on RL techniques with a focus on actor-critic methods, while section \ref{section:modeling} outlines our modelling of the state-action space and details our proposed methodology. In section \ref{section:toymarkets} we introduce some illustrative toy examples to showcase how the RL agent adapts in environments of varying complexities. This is followed by the numerical results and their analysis in section \ref{section:results_analysis}.
	\section{VOLATILITY FITTING}\label{section:vol_fitting_problem}
	
	Many methods exist in the industry to encode the implied volatility information extracted from the market. 
	For example, some practitioners can use closed or semi-closed parametric forms to define the variance (or the volatility)
	at every strike (see for example \cite{Valer2022volatility}). Other approaches consist in modeling the implied density 
	of the asset directly (see \cite{Gatheral2012arbitrage}) or even use a diffusion style model to fit the market at every expiry.\\
	For a survey of methodologies for constructing implied volatility surface (IVS) in practice, the reader can refer to \cite{IVS2011Cristian}.
	\\ 
	In this paper, we restrict ourselves to the modeling of the variance using a parametric function. 
	The techniques developed to solve the calibration of the volatility surface can be perfectly scaled to all the other parametrization choices.  
	Parametrization of the variance can take small to very large sets of coefficients \footnote{Some market providers use parametrizations with up to 20 coefficients.}. To keep the action space small, we restrict ourselves to 3 parameters and consider extending 
	to a larger set in a future study. Using 3 parameters is often not enough to fit the market properly, as there are not enough degrees of 	freedom to sufficiently bend the model implied volatility, this is why we will be adding to the RL results a benchmark generated from a
	classical optimizer.\\
	\\
	Traditionally, once the vehicle encoding the implied volatility is selected, the choice of the target is defined. In most cases, practitioners 
	aim at positioning the model implied volatility at the mid \footnote{The mid is computed from the aggregated bid/ask volatilities.} factoring 
	into account several effects such as the presence of arbitrage, smoothness of the term-structure, stability between consecutive fits and so on. 
	A careful selection of the reward is thus important when translating the classical model-based approach into a model-free reinforcement learning architecture. The fitting logic itself is driven by a set of deterministic rules where market data is massaged to generate aggregated quotes, the parametrization coefficients adjusted to approach the mid volatilities for some strikes and the output coefficients set cleaned from any bias induced by the numerical method.\\
	\\
	A first and natural alternative would be to consider the volatility fitting problem as a predictive exercise where the input $x$ is
	the collection of all available information at time $t$ and the output $\Delta y$ is the variation of the parametrization's parameters between $t$ and $t+\epsilon$. In such setting, the goal is to approximate the application $\phi$ such that $\Delta y = \phi \left(x\right)$. A very famous family of approximators are the feed forward neural networks which can learn large set of functions. While it is a viable and intuitive approach, solving the problem with such design has several downsides: on the one hand, the training set used to calibrate the weights of the neural networks captures solely a finite number of market regimes. This would require either a frequent re-calibration or training from the user to track changes in market regimes or an immense dataset to expand the coverage of the training set. On the other hand, labeling all the inputs $x$ assumes implicitly a choice of the objective function (for example low error to the mid). This means the calibrated neural network cannot be re-used when the user has a different objective function in mind \footnote{The user could build different neural nets to overcome this limitation but it also means having distinct training sets which could bare time and monetary cost challenges.} . \\
	\\
	The second alternative, which we introduce in this paper, is to swap the \textit{''deterministic algorithms''} with the Deep Reinforcement Learning framework. On the one hand, the actor neural network $\phi_{reward}$ provides a direct link between the inputs $x$ and the output $\Delta y$. On the other hand, the replay buffer is a dynamic training set which 
	can swiftly track the market evolution. Furthermore, when the objective function considered for the fitting is complex (e.g: desk PnL) and the standard optimizer too heavy to run \footnote{Multiple repricing can be required in the optimization routines.}, the RL approach can learn how to interactively adjust the volatility surface through intelligent cues (the rewards) freely available (e.g: in risk systems). As we will showcase in the coming sections, the Deep Reinforcement Learning (DRL) techniques are effective and can produce the same results as standard optimizers. They can be trained offline (for example outside market hours) or online after an initial warm-up phase. \\
	
	
	\vspace{-1mm}
	\section{PRELIMINARIES}\label{section:basics_rl}
	\subsection{Basics On Reinforcement Learning}
	
	Reinforcement Learning refers to a goal-directed learning and decision-making process where an agent acting within an evolving system, learns to make optimal decisions through repeated experiences gained by interacting with the environment without relying on supervision or complete model. (See Bertsekas(2005) \cite{bertsekas2005dynamic}, Sutton \& Barto(2018) \cite{ sutton2018reinforcement}).
	
	We consider a stochastic environment $ \mathcal{E} $, assumed fully-observed and thus formulate the problem of decision making of an agent acting within an environment over a finite time horizon, as a Markov Decision Process   $ (\mathcal{S},\mathcal{A},r,p) $, where at each time, the agent observes a current state $ s_{t} $ from the continuous state space $ \mathcal{S} \subset \mathbb{R}^{d}  $,  takes an action $ a_{t}  $ based on that current state from the continuous action space  $ \mathcal{A} \subset \mathbb{R}^{n}  $ and receives a scalar reward $ r_{t}  $ denoting numerical feedback from the stochastic scalar-valued reward function $ r =r(a_{t},s_{t} ) $ .
	
	The transition dynamics is given by $ p( s_{t+1} |s_{t} , a_{t})  $ and represents the probability density of the next state. The agent behaviour is defined by a policy $ \pi $ i.e. actions are drawn from $ \pi :\mathcal{S} \rightarrow  \mathcal{A} $ (resp. $ \mathcal{P}(\mathcal{A}) $) for a deterministic policy (resp. for  a randomized policy which maps the current state $ s_t $ to a probability distribution over the action space). 
	
	We will use $\rho_{\pi}(s_t)$ and $\rho_{\pi}(s_t,a_t)$ to denote the state and state-action marginals of the trajectory distribution induced by a policy $\pi$. 
	
	Denoting by $ \gamma   \in [0,1]$ a discounting factor, it is standard to define the return from a state as the sum of discounted future reward from following the policy $  \pi $:
	
	\begin{align}
	\label{eq:reward}
	R_{t} = R^{\pi}(s_t)  = \sum_{i =t}^{T} \gamma^{(i-t)}r(s_{i},a_{i} ).
	\end{align}
	Likewise, we define the state-action value function \footnote{ \ref{eq:value_function} and \ref{eq:value_func} provide a prediction of how good each state or each state/action pair is.} or Q-value as:

	\begin{align}
	\label{eq:value_func}
	Q^{\pi}(s, a)  &= \mathbb{E}^{\pi}[R_{t} | s_t=s,a_t=a] \\
	&= \sum_{i=t}^{T} \mathbb{E}_{(s_i,a_i)\sim \rho_\pi} [ \gamma^{(i-t)} r(s_{i},a_{i} )].
	\end{align}
	and the value or state value function:
	\begin{align}
	V^{\pi}(s)  = \mathbb{E}^{\pi}[R_{t} | s_t=s] = max_{a\in \mathcal{A} }Q^{\pi}(s, a).
	\label{eq:value_function}
	\end{align}
	
	The objective of Reinforcement Learning is to learn a policy $ \pi $ (i.e. a sequence of actions over time) which maximizes the expected sum of discounted reward from the system (value functions).
	In the case of Markov Decision Processes (MDP), an agent chooses its next action $ a_t = \pi(s_t) $
	according to its policy $ \pi $ depending on its current state $ s_t $; it
	then gets into a next state $  s_{t+1} = \mathcal{E} (s_t, a_t) $, depending on the response of the environment and gets the instantaneous reward $ r_t = r(s_t, a_t) $ \footnote{The functions $ \mathcal{E} $ , r and $ \pi $ may have random components.}.

	That agent reinforced with awards based on its interaction with the environment would learn to operate optimally in that environment to maximize the  cumulative rewards (trials and errors).
	
	The problem is thus formulated mathematically by defining the optimal value function for each state s as follow: 
	\begin{align}
	\label{eq:maxim}
	V^{*}(s)  = \sup_{\pi} V^{\pi}(s) = \sup_{\pi} \mathbb{E}^{\pi} \biggl[\sum_{i =t}^{T} \gamma^{(i-t)}r(s_{i},a_{i} ) \biggl| s_t=s \biggl ].
	\end{align}
	
	The standard approach to solve this optimization makes use of the well-known Dynamic Programming Principle (DPP) to derive the following variation of the recursive linear Bellman equation:
	
	\begin{align}
	\label{eq:NonLinearBellman}
	V^{*}(s)  = \sup_{a \in \mathcal{A}} \biggl[ \mathbb{E}[r(s,a)] + \gamma \mathbb{E}_{s'\sim p(s'|s,\pi(s))} \left[ V^{*}(s') \right] \biggl]
	\end{align}
	
	It follows from the literature that, in order to solve this problem, we can perform either value iteration \cite{watkins1992q}, policy iteration \cite{sutton2000policy,williams1992} or a combinasion of the two in actor-critic methods \cite{konda2000}.
	\subsection{Deep Reinforcement Learning Algorithms}
	Combining the above-mentioned ideas with Deep Learning has lead to significant performance improvement and the spreading of the so called Deep Reinforcement Learning (DRL).\\
	
	In fact, when we are in the setting of continuous, high dimensional action spaces, it is known from the literature \cite{recent2021advances} that classical tabular value-based and policy-based methods are challenging, we rather consider parametrized value functions $ Q(s, a; \theta^{Q})  $ and $ V (s, \theta^{V})  $  or policies $ (\pi(s, a; \theta^{\pi})) $  and then use Neural Networks as function approximators.\footnote{Thanks to the universal approximation theorem, Neural networks
	are known for their ability to approximate a wide range of non-linear functions in high
	dimension and to fit to many non-linear regression task \cite{Hornik1991,Cybenko1989,LBH2015}}. There are several popular neural network architectures amongst which the  fully-connected neural networks used in this study and for which the reader is invited to refer to the \textbf{appendix} \ref{appendix:subsection:FCNN} and \ref{appendix:subsection:FCNN_train} for more details.\\
	
	The combination of value-based and policy-based methods leads to Actor-Critic algorithms. This class of RL algorithms alternates between policy evaluation by computing the value function for a policy and policy improvement by using the value function to obtain a better policy. It can be extended to neural Actor-Critic algorithms by using neural networks for functional approximations. Deep Deterministic Policy Gradient (DDPG) and Soft Actor-Critic (SAC) algorithms are part of that class of RL algorithms which have demonstrated state-of-art performance on a range of challenging decision-making and control tasks. Those algorithms use a replay buffer to improve the performance of neural networks (see the \ref{appendix:subsection:ReplayMem} for more details).

	\paragraph{Deep Deterministic Policy Gradient (DDPG):} It is a model-free off-policy Actor-Critic algorithm, first introduced in \cite{lillicrap2015continuous}, which
	combines the Deep Q-Network (DQN) \cite{mnih2015human} and Deep Policy Gradient (DPG) algorithms. It extends the DQN to continuous action spaces by incorporating DPG to learn a deterministic strategy $ \pi^{D}  $. The critic estimates the $Q$-value function using off-policy data and the recursive Bellman equation:
	$${Q(s_t, a_t) = r(s_t, a_t) + \gamma Q\left(s_{t+1}, \pi^{D} (s_{t+1},\theta^{\pi}) \right)},$$
	
	The actor is trained to maximize the critic's estimated $Q$-values by back-propagating through both networks.
	
	To encourage exploration, it uses the following action (stochastic policy):
	\begin{align}
	\label{eq:ddpg}
	a_t \sim  \pi^{D} (s_t,\theta^{\pi}) + \mathcal{N}
	\end{align}
	where $ \mathcal{N} $ is either an Ornstein-Uhlenbeck (Ornstein and Uhlenbeck, 1930) noise process $ \mathcal{N} = \text{OU}(0, \sigma^2)$ \footnote{$ x_t= x_{t-1} + \theta(\mu-x_{t-1})dt +randn(size(\mu))\sigma\sqrt{dt}$.} (correlated)  or an additive Gaussian noise $ \mathcal{N} = \mathcal{N}(0, \sigma^2 I) $ (uncorrelated and time-independent), chosen according to the environment\footnote{We will used an additive Gaussian noise with a decreasing standard deviation to dampen exploration as the agent is learning the optimal policy.}.
	It thus treats the problem of exploration independently from the learning algorithm.
	
	\begin{figure}[H]
		\centering
		\includegraphics[width=0.5\textwidth]{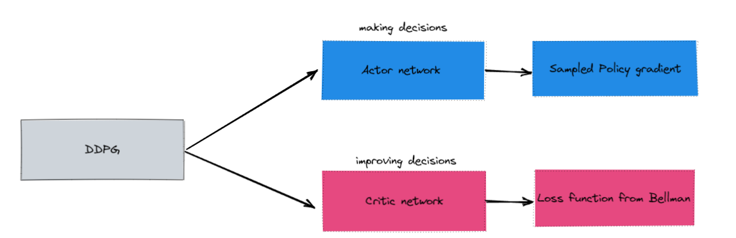}
		\caption{\label{fig:ddpg_algorithm} DDPG Framework. 
		}
	\end{figure}
	
	However, this algorithm is known to suffer from high sample complexity and sensibility to hyper-parameters \footnote{We witnessed it by changing differents hyperparameters.}. Subsequently, we also consider an extension known as SAC.

	\paragraph{Soft Actor Critic (SAC):}
	It is an off-policy actor-critic deep RL algorithm based on the maximum entropy reinforcement learning framework. It was introduced in  \cite{haarnoja2018soft} and extends the DDPG setting by allowing the actor to maximize the expected reward while also maximizing the entropy. The off-policy formulation enables reuse of previously collected data for efficiency, and the entropy maximization enables stability and provides a substantial improvement in exploration and robustness.  
	It favors stochastic policies by considering a more general and augmented entropy objective:
	
	\begin{align}
	\label{eq:maxent_objective}
	J_{\pi}  = \sum_{i=t}^{T} \mathbb{E}_{(s_i,a_i)\sim \rho_\pi} [ \gamma^{(i-t)} ( r(s_{i},a_{i} ) + \alpha\mathcal{H}(\pi(\cdot|s_i) )]. 
	\end{align}
	
	\begin{align*}
	\label{eq:maxent_objective2}
	\pi^{*} = \arg\max_{\pi \in \Pi} J_{\pi}.
	\end{align*}
	
	where $ \mathcal{H} $ is the entropy functional, $ \Pi $ a  tractable family of distributions and $\alpha$  is the temperature parameter which determines the relative importance of the entropy term against the reward, and thus controls the stochasticity of the optimal policy. 
	 
	 We want to project the improved policy into the desired set of policies in order to account for the constraint that $\pi \in \Pi$. We will rather use the information projection defined in terms of the Kullback-Leibler divergence and redefined the actor loss function \footnote{We choose KL divergence to project the improved policy into $ \Pi $. In principle we could choose others projections among which the Wasserstein projection.}.
	 
	 Let's denote by $Q_\theta$ the critic network, parameterized by $\theta$ and $\pi_\phi$ the actor network, parameterized by $\phi$ which outputs a probability distribution over the action space i.e. the best action to take from state $s_t$ is then sampled from the probability distribution $\pi_\phi(\cdot|s_t)$.
	 
	 We redefined the policy loss as the KL divergence or the gap between the probability distribution over the action space proposed by the actor network and the probability distribution induced by the exponentiated current Q function normalized by the factor $Z_{\theta}$ \footnote{The policy's density has the form of the Boltzmann distribution, where the Q-function serves as the negative energy. It can be show that it is an optimal solution for the maximum-entropy RL objective $ \ref{eq:maxent_objective} $. }:

	  In fact, the high values of the latter indicates the areas of the action space where the cumulative expected sum of rewards is approximated to be high, minimizing the KL divergence means getting an efficient actor network associated to actions yielding highly rewarded trajectories. 
	 
	 With the definition of the KL divergence, we have:
	
	\begin{align*} 
	J_{\pi}(\phi) &=  \EE_{s_t \sim \mathcal{D}} \left[ D_{KL}\enstq{\pi_{\phi}(\cdot|s_t)}{\frac{\exp\left(\frac{1}{\alpha} Q_{\theta}(s_t, \cdot)\right)}{Z_{\theta}(s_t)}}\right] 
	\end{align*}
	
	\begin{align*} 
	&= \EE_{s_t \sim \mathcal{D}} \left[ \int_{a_t} \pi_{\phi}(a_t|s_t) \log \frac{\pi_{\phi}(a_t|s_t)}{\exp\left(\frac{1}{\alpha}Q_{\theta}(s_t, a_t)\right)/Z_{\theta}(s_t)} da_t \right] \\
	&= \EE_{s_t \sim \mathcal{D}} \left[ \EE_{a_t \sim \pi_{\phi(\cdot|s_t)}}[\alpha \log \pi_{\phi}(a_t|s_t) - Q_{\theta}(s_t, a_t)] \right] + C^{\text{ste}} 
	\end{align*}
	
	We use a re-parametrization trick for the probability law according to which actions are
	drawn: \footnote{Instead of the standard likelihood ratio method. This may result in a lower variance estimator and still is an open question on which no consensus is found in the literature.} i.e.
	
	\begin{align}
	a_t = f_\phi(\epsilon_t; s_t) =  \mu_{\phi}(s_t) + \epsilon_t \sigma_{\phi}(s_t),  
	\end{align}
	where $\epsilon_t$ is an input noise vector such as a spherical Gaussian ($\epsilon_t \sim \mathcal{N} (0, I)$ )
	
	Choosing the optimal temperature is non-trivial since it needs to be tuned for each task. We overcome this issue as proposed in \cite{haarnoja2019sac} by formulating a different maximum entropy RL objective, where the entropy is treated as a constraint.
	
	 Our aim is to find a stochastic policy with maximal expected return that satisfies a minimum expected entropy constraint. Formally, we need to solve the entropy-constrained maximum expected return objective  problem. 
	 
	\begin{align*}
	\max_{\pi_{t:T}} \E{\rho_\pi}{\sum_{i=t}^T  \gamma^{(i-t)} r(s_i, a_i)} 
	\\ \text{ s.t. } \E{(s_t, a_t)\sim\rho_\pi}{-\log(\pi_t(a_t|s_t))} \geq \bar{\mathcal{H}}\ \ \forall t
	\end{align*}
	where $\bar{\mathcal{H}}$ is a  predefined or desired minimum policy expected entropy threshold \footnote{we choose the entropy target $ \bar{\mathcal{H}} $ to be -dim($ \mathcal{A} $) as proposed by the article.}.
	
	 We use the Lagrangian relaxation\cite{Lagrangianduality2010} to transform the entropy constraint into a penalty, yielding an expression that is easier to solve.

	We solve the optimal dual variable (also known as “Lagrange multiplier”) at different steps backward in time, $\alpha_t^*$ (the optimal temperature at step t) after solving for $Q^*_t$ and $\pi_t^*$:
	
	By repeating this process, we can learn the optimal temperature parameter in every step by minimizing the same objective function:
	\begin{align}
	\alpha_t^* = \arg \min_{\alpha_t} \E{a_t\sim \pi_t^*}{- \alpha_t\log\pi_t^*(a_t|s_t; \alpha_t) - \alpha_t \bar{\mathcal{H}}}.
	\label{eq:optimal_temperature}
	\end{align}
	\\

	In the finance industry, new ideas coming from reinforcement learning have been
	developed and adapted to financial problems with many successes. For surveys in the literature on RL applications in finance, the reader is invited to refer to \cite{recent2021advances, DeepRL2022Hedging,cao2021hedging,zhao2021high,ganesh2019reinforcement,spooner2020robust,Ye2020}.
	This has attracted a lot of attention in applying RL techniques to improve decision-making in various financial markets and thus has instigated the following natural extension to the volatility fitting problem.  
	
	\vspace{0.5cm}
	
	
	\vspace{-1mm}
	\section{MODELING SPECIFICATION}\label{section:modeling} 
	
	In this section, we frame the fitting problem into an RL setting, translating the classical definitions into the reinforcement
	learning framework. 
	We conclude this section by detailing the implementation of the Actor-critic algorithms 
	(DDPG and SAC) and explaining how they were adapted to the fitting game.

	\subsection{Problem Statement In Reinforcement Learning Framework }
	
	\paragraph{Parametrisation and market observables:}

	Let us fix a maturity T and observe the market at time $ t_i $, i.e. a collection $ \left\lbrace  \sigma^{Ask}_{Call/Put} (t_i, \kappa_{j}), \sigma^{Bid}_{Call/Put}(t_i, \kappa_{j}) \right\rbrace_{ j \in  \left[ 1,...,n \right]} $ of markets quotes for different moneyness $\left( \kappa_{1},...,\kappa_{n}\right)  $.
	
	\begin{figure}[H]
		\centering
		\includegraphics[width=0.48\textwidth]{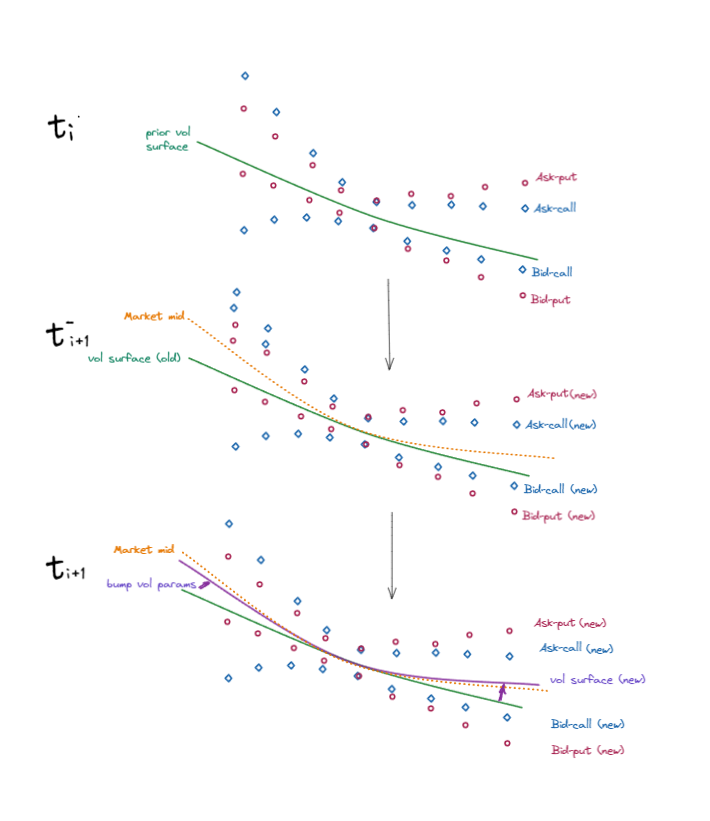}
		\caption{\label{fig:RL_agent} A snapshot of the agent acting in the market: The old volatility surface is bumped to a new one following the move of the market, conditional on prior volatility surface. 
		}
	\end{figure}

	We consider a setting with a parametrization $ \Psi^{vol}_{\vec{\theta}} $ \footnote{We can think about the Stochastic volatility inspired (SVI) parameterization of the total implied variance for a fixed time to maturity proposed by Gatheral \cite{SVI2004Gatheral, Gatheral2012arbitrage, GeneralizedSVI2016} . \[\Psi^{SVI}_{\vec{\theta}}(k)^{2} = a+b(\rho(k-m)+\sqrt{(k-m)^2+\sigma^2})\]
		where $k$ is the log-forward moneyness, and $\vec{\theta} = (a,b,\rho,m,\sigma)$.} of the volatility slice with a vector of K parameters denoted by $  \vec{\theta}_{t_i}  = \left( \theta_{t_i}^{1},...,\theta_{t_i}^{K}\right)$.
	The market is moving over time, and at time $ t_{i+1} $, this fitting problem consists of finding the optimal vector $ \vec{\theta}_{t_{i+1}}^{*} $ which maximizes an objective function. The parametrisation $ \Psi^{vol}_{\theta} $ is classically optimized to be close to the mid volatilities $\left\lbrace \sigma^{Mid}(t_i, \kappa_{j}) \right\rbrace_{ j \in  \left[ 1,...,n \right]}  $ with $ \sigma^{Mid}(t_i, \kappa_{j}) = \frac{\sigma^{Ask} (t_i, \kappa_{j}) + \sigma^{Bid} (t_i, \kappa_{j})}{2} $ (See Figure \ref{fig:RL_agent} ) .\\
	
	In the RL framework, we leave the problem open to more than just matching the mid as much as possible. In fact, our goal is to maximize a certain reward defined in term of an objective function: with a market move at time $ t_{i+1} $, the challenge for an agent is to bump the parameters of our volatility slice parametrization $ \Psi^{vol}_{\vec{\theta}} $ in order to maximize a certain reward.\\
		
	Subsequently, for a range of moneyness, and starting from a prior parametrized volatility surface $\Psi^{vol}_{\vec{\theta}_{t_i}} $ ( i.e. the latest fit is the prior input), the problem is to find the shifted vector $ \Delta\vec{\theta}_{t_i} = \left( \Delta \theta_{t_i}^{1},...,\Delta\theta_{t_i}^{K}\right) $ to bump the old parameters $ \vec{\theta}_{t_i} $ in order to maximize our selected reward function.
		
	\paragraph{State and Action spaces:}
	Our state is the set of tuple consisting of the observed markets quotes and the prior volatility parameters: $$s_{t_i} = \left( \sigma^{Bid}(t_i, \kappa_{j}),\sigma^{Ask}(t_i, \kappa_{j}), \theta_{t_{i-1}}^{1},...,\theta_{t_{i-1}}^{K}\ \right)_{j \in [1,n]} \in \mathbb{R}^{N}$$ (with $ N \ge  2n+K $), where we consider the bid and ask market implied volatility, i.e. the state space is continuous  \footnote{We may also rather consider  $\left( \sigma^{Mid}(t_i, \kappa_{j}), \sigma^{\text{spread}}(t_i, \kappa_{j}), \theta_{t_{i-1}}^{1},...,\theta_{t_{i-1}}^{K}\ \right)_j \in \mathbb{R}^{N}$ (with $ N \ge  2n+K $), where $ \sigma^{\text{spread}}(t_i, \kappa_{j}) := \sigma^{Ask} (t_i, \kappa_{j}) - \sigma^{Bid} (t_i, \kappa_{j}) $ is the bid-ask volatility spread.}.

	From the RL perspective, the fitting cycle looks as follows: At time $ t_{i+1}^{-} $,
	
	\begin{enumerate}
		\item The agent sees the state $ s_{t_{i+1}} $ of our environment built on the market bid and ask volatilities (homogeneous variable) and the old volatility surface parameters $ \vec{\theta}_{t_i} $ (endogenous variable),
		\item It takes some action $ a_{t_{i+1}} = \Delta\vec{\theta}_{t_{i+1}} = \theta_{t_{i+1}} - \theta_{t_{i}} $ on the adjustment of the old parameters $ \vec{\theta}_{t_i} $, and receives some reward $ r(s_{t_{i+1}},a_{t_{i+1}} = \Delta\vec{\theta}_{t_{i+1}})$.

		\item The resulting volatility slice maximizes the reward and it boils down a new state $ s_{t_{i+2}} $ for the next move. 

	\end{enumerate}
	
	The error term between our volatility surface and the mid market is denoted $ \xi (\vec{\theta}_{t_{i+1}} ) $ . 
	
	Since the vector of bumps $ \Delta\vec{\theta} $ ( resp. the actions) can take any values in $ \mathbb{R}^{K} $ ( resp. $ \mathbb{R}^{K} $) space, it draws down a continuous actions spaces.

	\paragraph{Reward functions:}
	We use two metrics to measure the goodness of the fit: \footnote{We can also design a reward functional taking into account a set of constrains like the liquidity, macro-events, term-structure, stability.}
	\begin{itemize}
		\item Mean squared error (MSE)  defined by : 
		\begin{align}
		\label{eq:mse}
		\xi (\vec{\theta}_{t_i} ) := \sum_{j = 1}^{n } (\sigma^{Mid}(t_i, \kappa_{j}) -  \Psi^{vol}_{\vec{\theta}_{t_i}}(k_j) )^{2}
		\end{align}
		It is used to penalise larger deviations from the true values, making it an effective metric for assessing the quality of fit.
		
		\item Black Scholes Vega weighted Mean squared error (BMSE)  defined by :
		\begin{align}
		\label{eq:vega_mse}
		\xi (\vec{\theta}_{t_i} )  := \sum_{j = 1}^{n } \text{vega}_{BS}(\kappa_{j}) ( \sigma^{Mid}(t_i, \kappa_{j}) -  \Psi^{vol}_{\vec{\theta}_{t_i}}(\kappa_{j}) )^{2}
		\end{align}
		It is used to penalise markets quotes located at the neighbourhood of the at-the-money region.

	\end{itemize}
 $ \mathbf{Remark:} $ We can also use the Scaled Mean squared error (SMSE) 
  \footnote{ Setting $ \sigma^{\text{spread}}(t_i, \kappa_{j}):= \sigma^{Ask}_{t_i, \kappa_{j}} - \sigma^{Bid}_{t_i, \kappa_{j}} $ the volatility spread, we define the SMSE :
		\begin{align}
		\label{eq:Scaled_mse}
		\xi (\vec{\theta}_{t_i} )  := \sum_{j = 1}^{n} (\frac{ \sigma^{Mid}(t_i, \kappa_{j}) -  \Psi^{vol}_{\vec{\theta}_{t_i}}(\kappa_{j}) }{\sigma^{\text{spread}}(t_i, \kappa_{j})} )^{2}
		\end{align}
	 } to give more importance to market quotes with small bid-ask spread.
	
	Within all these cases, we defined the reward as the opposite of that error term: 
	
	$ r(s_{t_{i}},a_{t_{i}} = \Delta\vec{\theta}_{t_i}) = - \xi (\vec{\theta}_{t_{i}} )  $

	\subsection{Algorithms Overview For Volatility Fitting}
	
	In this setting, the state-action value can be written:
	\begin{align}
	  \label{eq:value_vol}
	  Q^{\pi}(s_{t_{i}}, a_{t_{i}} ) =  - \sum_{k=t_i}^{T} \mathbb{E}_{(s_{k},a_{k})\sim \rho_\pi} [ \gamma^{(k-t_i)} \xi (\vec{\theta}_{k} )]
	\end{align}
	
	We adapt the framework of DRL technique  and provide a natural extension of DDPG and SAC to the volatility fitting problem, and adjust our exploration preference (decaying the standard deviation  of the exploration noise and the temperature coefficient with steps as we draw close to the optimal reward). We also choose in the $\textit{static and sequential scenarios}$ to store only transitions which improve the replay memory (i.e. for which the reward is greater that the worst reward in the replay buffer\footnote{see \ref{staticMarket} and \ref{sequentialMarket}. In the dynamic scenario, we just remove the oldest transition in the replay memory}).  
	
	Note that the sampled minibatch from the replay buffer in our algorithms  is  a state-action-reward-next state tuple of the form
	$$(s_{t_1},{a}_{t_1},r_{t_1},s_{{t_2}}),\dots,(s_{t_K},{a}_{t_K},r_{t_K},s_{t_K+1}).$$
	
	Using the observed markets quotes and the prior volatility parameters to represent the market state present the challenge of feature scaling regarding their order of magnitude. In this context, unlike supervised learning, we cannot simply re-scaled the features in the naive way since the a-priori distribution of states under the optimal policy is unknown. We used batch normalization (dynamic feature scaling or running means and standard deviation) of states as solution to this problem. 
	
	The critic and actor networks are then updated according to these samples with a normalization of layer inputs as proposed in \cite{batchNorm} to reduce the internal co-variate shift. The reader should refer to the appendix \ref{appendix:section:Hyperparams} for the hyper-parameters used for the training.
	
	\paragraph{Deep Deterministic Policy Gradient (DDPG):}
	In our setting (see section \ref{section:toymarkets} ),  we perform exploration using a Gaussian Noise (GN) instead of Ornstein
	Uhlenbeck (OU) action noise  as GN is time-independent while OU noise presents some auto-correlation.\\
	
	\begin{figure}[H]
		\centering
		\includegraphics[width=0.38\textwidth]{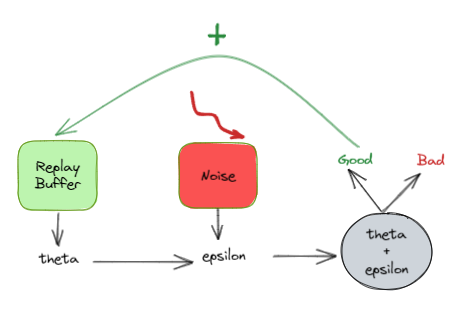}
		\caption{\label{fig:gn_ddpg_algorithm} Replay Memory and Noise in the DDPG framework in the static and sequential toy problems.}
	\end{figure}

	We choose to dampen exploration by powerly decaying the standard deviation of the Gaussian Noise (Fig. \ref{fig:gn_ddpg_algorithm} ).
	
	 \begin{algorithm}[H]
		\caption{\textbf{DDPG variant for Volatility fitting}}
		\label{alg:DDPG}
		\begin{algorithmic} [1]
			\STATE \textbf{Input}: Randomly initialize an actor network $\pi^D(s;\theta)$, a critic network $Q(s,a;\phi)$,with parameters $\theta^{(0)}$ and $\phi^{(0)}$
			\STATE  Initialize target networks $ \bar{\pi}^D $ and $ \bar{Q} $ with parameters $\bar{\phi}^{(0)}\leftarrow \phi^{(0)}$ and $\bar{\theta}^{(0)}\leftarrow \theta^{(0)}$, 
			\STATE Initialize the replay buffer $\mathcal{RB}$  
			\FOR{$n=0,\ldots,N-1$}
			\STATE Initialize a gaussian process $\mathcal{N}$ for action exploration and fix LearningFlag= True. 
			\STATE Initialize state $s_0$ with a flat volatility parameters
			
			\FOR{$t=1,\ldots,M$} 
			\STATE Select the bumps  $a_t\sim\pi^D(s_t;\theta^{(n)})+\epsilon_t$  with $\epsilon_t\sim\mathcal{N}(0,\sigma_n)$, $ \sigma_n = \text{max}(\sigma_0(1-\frac{n}{N} )^{4},\sigma_{\text{min}}) $
			\STATE Execute the bumps $ \pi^D(s_t;\theta^{(n)}) $ (deterministic) and $a_t$(exploration), then receive reward $r^D_t$, $r_t$ and observe new state $s_{t+1}$
			\STATE Smartly Store the transition $(s_t,a_t,r_t,s_{t+1})$ in $\mathcal{RB}$ , (see Fig.\ref{fig:gn_ddpg_algorithm})
			
			\IF{LearningFlag} 
			\STATE Sample a random mini-batch of $N_{batch}$ transitions $\{(s_{(i)},a_{(i)},r_{(i)},s_{(i+1)})\}_{i=1}^{N_{batch}}$ from $\mathcal{RB}$
			\STATE Compute the target \\ $Y_i = r_i+\gamma \bar{Q}(s_{i+1},\bar{\pi}^D(s_{i+1};\bar{\theta}^{(n)});\bar{\phi}^{(n)})$.
			\STATE Update the critic by minimizing the loss: $\phi^{(n+1)}=\phi^{(n)}-\beta\nabla_{\phi}\mathcal{L}_{\rm DDPG}(\phi^{(n)})$
			\STATE Update the actor by using the sampled policy gradient: $ \theta^{(n+1)}=\theta^{(n)}+\alpha\nabla_{\theta}J(\theta^{(n)})$ 
			\STATE Update the target networks via polyak averaging:
			
			\begin{eqnarray*}
				\bar{\phi}^{(n+1)}&\leftarrow& \tau \phi^{(n+1)} + (1-\tau) \bar{\phi}^{(n)}\\
				\bar{\theta}^{(n+1)}&\leftarrow& \tau \theta^{(n+1)} + (1-\tau) \bar{\theta}^{(n)}
			\end{eqnarray*}
			
			\ENDIF
			\STATE Update LearningFlag wrt to a preset criterion\footnotemark
			\ENDFOR
			\ENDFOR
		\end{algorithmic}
	\end{algorithm}
	\footnotetext{$r^D_t > \mathcal{R}_0 $ (in the static and sequential toy case) or $r^{m}_t > \mathcal{R}_0 $(in the dynamic case) where $\mathcal{R}_0$ is a reward threshold and $ r^{m}_t $ the mean reward of the evaluation episode(see \ref{dynamicMarket} ). \label{criteria}}
	
	$\nabla_{\theta}J(\theta)\approx \frac{1}{N_{batch}}\sum_{i}\nabla_a Q(s,a;\phi)\nabla_{\theta}\pi^D(s;\theta)|_{s=s_i,a=a_i} $ \footnote{this analytical gradient is useless in practice with common deep learning frameworks such as pytorch
		since these libraries are such that writing the policy loss
		is enough: the gradient is automatically computed when the backward pass is performed.} 
	
	$\mathcal{L}_{\rm DDPG}(\phi)=\frac{1}{N_{batch}}\sum_{i=1}^{N_{batch}} (Y_i-Q(s_i,a_i;\phi))^2$
	
	\paragraph{Soft Actor Critic (SAC):} 
	\vspace{25px}
	We use two soft Q-function to mitigate positive bias in the policy improvement step.
	We also learn the optimal temperature coefficient at each step by minimizing the dual objective in Equation $ \ref{eq:optimal_temperature} $
	
	\begin{algorithm}[H]
		\caption{\textbf{SAC variant for Volatility Fitting}}
		\label{alg:soft_actor_critic}
		\begin{algorithmic} [1]
			\STATE \textbf{Input}: Randomly initialize an actor network $\pi(s;\phi)$, two critics networks $Q(s,a;\theta_1)$ and $Q(s,a;\theta_2)$, with parameters $\phi^{(0)}$,  $\theta_1^{(0)}$ and $\theta_2^{(0)}$. \\ Initialize $\alpha^{(0)}$ and set $ \bar{\mathcal{H}}= - \text{dim}( \mathcal{R}^{K}) = -K $
			\STATE  Initialize target networks $ \bar{Q} $ with weights \\ $\bar{\theta}_1^{(0)}\leftarrow \theta_1^{(0)}$ and $\bar{\theta}_2^{(0)}\leftarrow \theta_2^{(0)}$, 
			\STATE Initialize an empty replay pool \\ $\mathcal{RB}\leftarrow\emptyset$ and fix LearningFlag= True.
			\FOR{$n=0,\ldots,N-1$} 
			\STATE Initialize state $s_0$ with a flat volatility parameters
			\FOR{$t=1,\ldots,M$} 
			\STATE Sample the vector of bumps from the policy  \\ $a_t \sim \pi_\phi(a_t|s_t) := \mu_{\phi}(s_t) + \epsilon_t \sigma_{\phi}(s_t)$ 
			\STATE Execute the bumps $ \mu_{\phi}(s_t) $ (deterministic) and $a_t$(exploration) , then receive rewards $r^D_t$, $r_t$ and observe new state $s_{t+1} \sim p (s_{t+1}| s_t, a_t)$
			
			\STATE Store the transition in the replay pool \\ $\mathcal{RB} \leftarrow \mathcal{RB} \cup \left\{(s_t, a_t, r(s_t, a_t), s_{t+1})\right\}$ 
			\IF{LearningFlag}
			\STATE Sample a random mini-batch of $N_{batch}$ transitions $\{(s_{(i)},a_{(i)},r_{(i)},s_{(i+1)})\}_{i=1}^{N_{batch}}$ from $\mathcal{RB}$
			\STATE Compute the target $y_i =$ \\ $r_i+\gamma \left( \bar{Q}_{\theta}(s_{i+1}, a_{i+1}) - \alpha^{(n)} \log \pi (a_i|s_i;\phi^{(n)}) \right) $. 
			
			\STATE Update the Q-function parameters: \footnotemark for $j\in\{1, 2\}$  
			$\theta_j^{(n+1)} \leftarrow \theta_j^{(n)}  - \lambda_Q \hat \nabla_{\theta_j} J_{Q_j}(\theta_j^{(n)})$ 
			
			\STATE Update policy weights: \footnotemark
			\\ $\phi^{(n+1)} \leftarrow \phi^{(n)} - \lambda_\pi \hat \nabla_\phi J_\pi(\phi^{(n)})$  
			
			\STATE Adjust temperature: \\ $ \alpha^{(n+1)} \leftarrow \alpha^{(n)} - \lambda \hat \nabla_\alpha J(\alpha^{(n)})$ 
			\STATE Update the target networks weights via polyak averaging: \\
			$\bar\theta_j^{(n+1)}\leftarrow \tau \theta_j^{(n+1)} + (1-\tau)\bar\theta_j^{(n)}$ for $j\in\{1,2\}$
			\ENDIF
			\STATE Update LearningFlag wrt to a criterion\footnotemark
			\ENDFOR
			\ENDFOR
			\ENSURE Optimized parameters $\theta_1^{*}$, $\theta_2^{*}$ and $\phi^{*}$
		\end{algorithmic}
	\end{algorithm}

	\footnotetext[24]{by minimizing the MSE loss against Bellman backup $ J_{Q_j}(\theta_j^{(n)}) = \frac{1}{N_{batch}}\sum_{i=1}^{N_{batch}} (y_i- \bar{Q}(s_i,a_i;\theta_j^{(n)}))^2 for j\in\{1, 2\} $ }
	
	\footnotetext[25]{by minimizing the Entropy-regularized policy loss:\\  $ J_\pi(\phi) =  \EE_{s_t \sim \mathcal{D}} \left[ \EE_{\epsilon \sim \mathcal{N}(0,I)}[ \alpha \log \pi_{\phi}(f_{\phi}(\epsilon;s_t)|s_t) - Q_{\theta}(s_t, f_{\phi}(\epsilon;s_t))] \right]
		\\  
		\quad Q_\theta(s_i,a_i) = \min (Q(s_i,a_i;\theta_1^{(n)}) , Q(s_i,a_i;\theta_2^{(n)})).$} 
	\footnotetext[26]{same as \ref{criteria} }

	
	\vspace{1cm}
	
	\section{INSIGHTS FROM TOY-MARKETS}\label{section:toymarkets}
	In order to assess the performance of the reinforcement learning algorithms for the volatility fitting problem, we
	consider toy markets of increasing complexity. The data used for the toy markets is generated synthetically to reflect real market configurations. As we will detail in the coming lines, we start with a simple scenario where market data is static. We then relax this hypothesis to slowly account for the market dynamics.   
	\subsection{Static Scenario}\label{staticMarket}
	In this scenario, the market quotes are static and do not change during the full experiment. In particular, the initial state is a flat volatility surface and the fitting episode ends after one step. The state space is degenerated into a single state	but the number of actions is infinite. The goal of the agent is to immediately detect the right parametrization bumps to apply in order to approach the market mid. 
	\subsection{Sequential Scenario}\label{sequentialMarket}
	In this scenario, the market quotes are static and do not change during the full experiment. In particular, the initial state is a flat volatility surface and the fitting episode ends after several steps (e.g: 50 steps \footnote{This order of magnitude is in line with the number of fits that can happen between the open and the close if a fit is scheduled every
	10mn.} ). Both the state and action spaces are infinite. The goal of the agent is to  detect the right parametrization bumps to apply in order to approach the market mid in an episode. Due to the effect of the discount factor, the agent is encouraged to detect the right parameter shifts from the first step. 
	
	\subsection{Quasi-Dynamic Scenario}\label{dynamicMarket}
	In this scenario, we allow the market to evolve freely during a full episode. In particular, the mids and spreads (for different moneyness) are random quantities with distinct marginals and a common joint distribution\footnote{The marginals are assimilated to normal distributions and the joint-dynamic is treated as a Gaussian copula}. The means, variances and correlations are calibrated using $\textit{''real-market''}$ data for a full trading day. This scenario is a clear cut versus the static and sequential market as: 
	\begin{itemize}
		\item The dimensionality of the state space is higher as the market quotes components is changing at every step. 
		\item The success criteria is more subtle as it doesn't only consist in approaching a single static terminal curve but to have the agent approach and track the market 	for several steps.
	\end{itemize} 
	In this setting, we perform three successive operations: 
	\begin{enumerate}
		\item Training Phase: we calibrate the hyper-parameters (e.g: learning rates, volatility noise, replay buffer size, etc) to increase the average reward in the evaluation episodes. The evaluation episodes are modes where all the randomness of the DRL algorithms is removed and no updates to the neural networks is performed. This training phase has a side benefit as it helps to determine a stopping threshold (for learning) for our agents.
		\item Validation Phase: in this phase, 	with the optimal configuration of hyper-parameters and the stopping criterion for learning, we train several agents and upon completion of training, we compare the performance of different successful agents under different seeds to select the most promising candidate. 
		\item Testing Phase: we assess the performance of the best agent in a test episode. 
	\end{enumerate}

	\section{RESULTS AND ANALYSIS}\label{section:results_analysis}
	In this section, we evaluate the performance obtained by the reinforcement learning algorithms \footnote{Note: The DDPG and SAC  are variants of the seminal algorithms adjusted for the volatility fitting problem.} for the toy problems described in \ref{section:toymarkets} . The algorithms performance is tested against different market configurations that can be encountered in live trading activity.
	
	\subsection{Static Market}
	We compare the performance of the DDPG and SAC (averaged across multiple random seeds) against a classical optimizer (Benchmark) in \ref{staticMarket}. 
	We first show a summary table with the final rewards for different market configurations before displaying the implied volatility 
	at convergence.

	\part*{ \large MSE Reward Summary:}

	As can be seen from table \ref{tab:title1} and \ref{tab:title2}, the final rewards achieved by the variants of the actor-critic algorithms is close to the one coming from the optimizer. The 
	performance is stable across the different market configurations considerered. 
	\vspace{1cm}
	\begin{minipage}{\linewidth}
		\centering
		\captionof{table}{DDPG MSE Rewards} \label{tab:title1}
		\begin{tabular}{lrrrr}
			\toprule {} type&      Skew &  High Smile &  Inv. Smile \\ \midrule Bench          & -0.011913 &    -0.000022 &      -0.000044 \\
			seeds' avg &  -0.012145 &     -0.000163 &       -0.000315 \\
			\bottomrule
		\end{tabular} 
	\end{minipage}
	\\\\
	\vspace{1cm}
	\begin{minipage}{\linewidth}
		\centering
		\captionof{table}{SAC MSE Rewards} \label{tab:title2}
		\begin{tabular}{lrrrr}
			\toprule {} type&      Skew &  High Smile &  Inv. Smile \\ \midrule Bench          & -0.011913 &    -0.000022 &      -0.000044 \\
			seeds' avg &  -0.011945 &     -0.000221 &       -0.001800  \\
			\bottomrule
		\end{tabular} 
	\end{minipage}

	\subsubsection{High Smile Market (MSE)}
	
	\begin{figure}[H]
		\centering
		\includegraphics[width=0.41\textwidth]{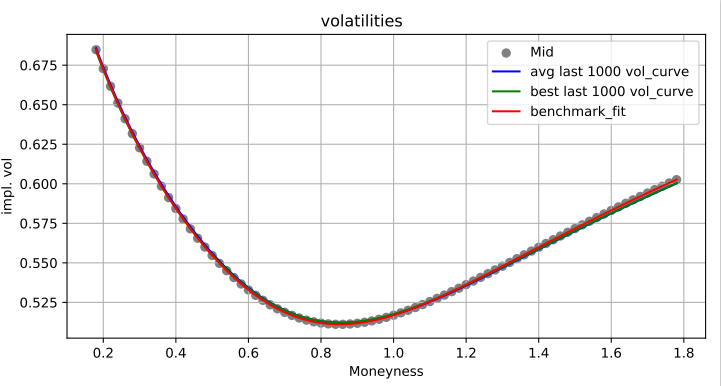}
		\captionof{figure}{\label{fig:smile}Implied volatility snapshot in a ''High Smile'' configuration with \textbf{DDPG}: A Monte-Carlo on 5 differents random seeds is performed with a power decaying noise. We represent in ($ \textcolor{green}{green} $) the best response of the agent amongst the last 1000 episodes and in ($ \textcolor{blue}{blue} $) the mean of the last 1000 volatility slices. 
		}
	
	\end{figure}

    \begin{figure}[!h]
    	\centering
    	\includegraphics[width=0.41\textwidth]{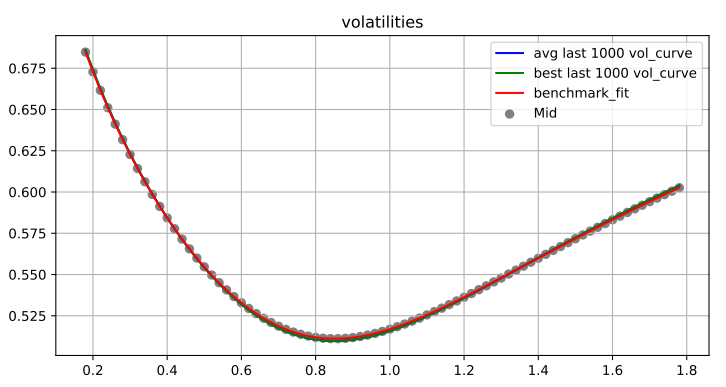}
    	\captionof{figure}{\label{fig:smile} Implied volatility snapshot in a ''High Smile'' configuration with \textbf{SAC}: A Monte-Carlo on 5 different random seeds is performed with automatic entropy adjustement. We represent in ($ \textcolor{green}{green} $) the best response of the agent amongst the last 1000 episodes and in ($ \textcolor{blue}{blue} $) the mean of the last 1000 volatility slices. 
    	}
    \end{figure}

	\subsubsection{Skew Market (MSE)}

	\begin{figure}[H]
		\centering
		\includegraphics[width=0.39\textwidth]{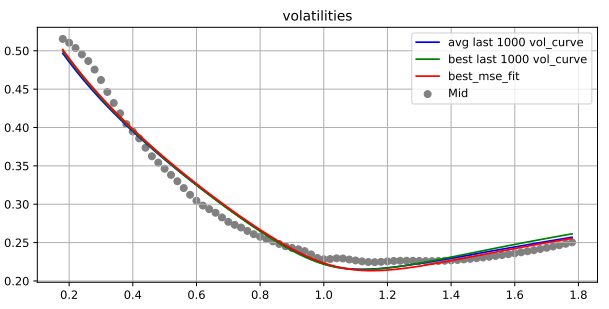}
		\caption{\label{fig:skew}Implied volatility snapshot in a ''Skewed'' configuration with \textbf{DDPG}: A Monte-Carlo on 5 differents random seeds is performed with a power decaying noise. We represent in ($ \textcolor{green}{green} $) the best response of the agent amongst the last 1000 episodes and in ($ \textcolor{blue}{blue} $) the mean of the last 1000 volatility slices.  }
	\end{figure}
	
	\begin{figure}[H]
		\centering
		\includegraphics[width=0.4\textwidth]{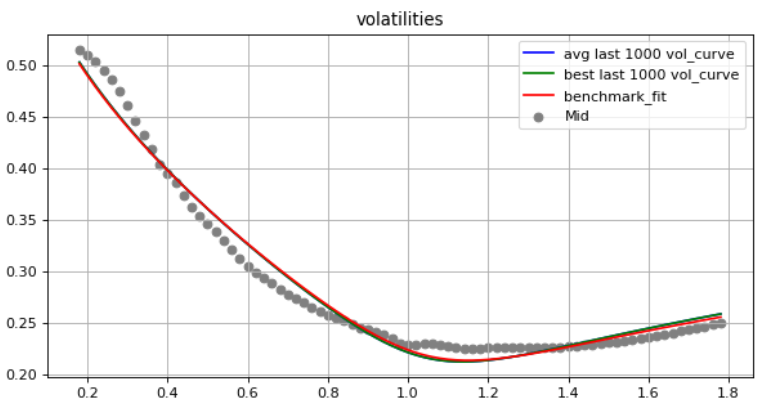}
		\captionof{figure}{\label{fig:smile} Implied volatility snapshot in a ''Skewed'' configuration with \textbf{SAC}: A Monte-Carlo on 5 different random seeds is performed with automatic entropy adjustement. We represent in ($ \textcolor{green}{green} $) the best response of the agent amongst the last 1000 episodes and in ($ \textcolor{blue}{blue} $) the mean of the last 1000 volatility slices. 
		}
	\end{figure}

	\subsubsection{Inverse Smile Market (MSE)}

	\begin{figure}[H]
		\centering
		\includegraphics[width=0.4\textwidth]{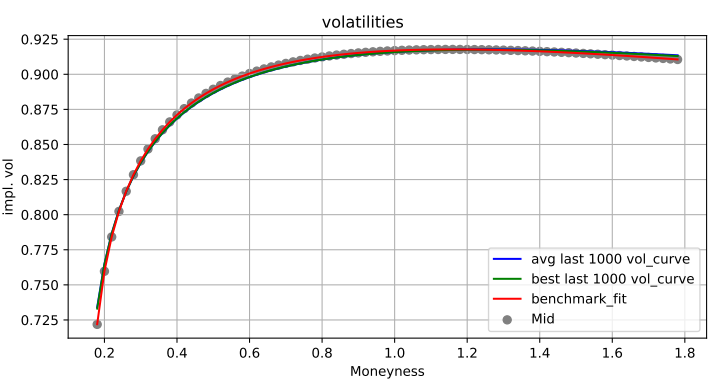}
		\caption{\label{fig:inversesmile} Implied volatility snapshot in a ''Inverse Smile'' configuration with \textbf{DDPG}: A Monte-Carlo on 5 differents random seeds is performed with a power decaying noise. We represent in ($ \textcolor{green}{green} $) the best response of the agent amongst the last 1000 episodes and in ($ \textcolor{blue}{blue} $) the mean of the last 1000 volatility slices. }
	\end{figure}

    \begin{figure}[H]
    	\centering
    	\includegraphics[width=0.44\textwidth]{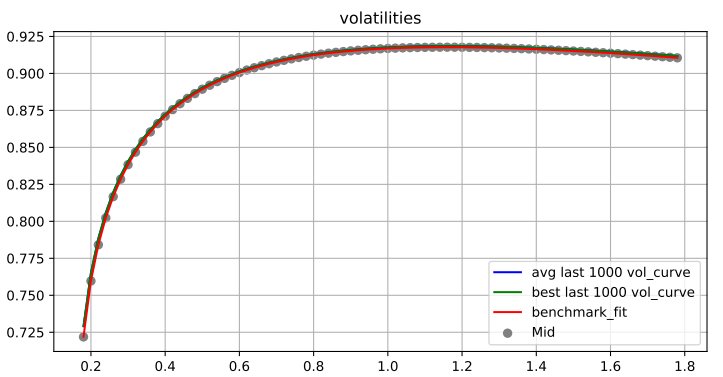}
    	\captionof{figure}{\label{fig:smile}Implied volatility snapshot in a ''Inverse Smile'' configuration with \textbf{SAC}: A Monte-Carlo on 5 different random seeds is performed with automatic entropy adjustement. We represent in ($ \textcolor{green}{green} $) the best response of the agent amongst the last 1000 episodes and in ($ \textcolor{blue}{blue} $) the mean of the last 1000 volatility slices.
    	}
    \end{figure}
	For more details on the black-scholes vega weighted reward, please refer to the \textbf{appendix} \ref{appendix:subsection:appendix1}
	
	\subsection{Sequential Scenario}
	We compare the performance of the DDPG and SAC (averaged across multiple random seeds) against a classical optimizer (Benchmark) in \ref{sequentialMarket}. 
	In this scenario, more than one step is allowed per episode. We first show a summary table with the final rewards for different market configurations
	before displaying the implied volatility at convergence.

	\vspace{1cm}
	\part*{ \large MSE Reward Summary:}
	As can be seen from table \ref{tab:title3} and \ref{tab:title4}, the final rewards achieved by the variants of the actor-critic algorithms is close to the one coming from the optimizer. The 
	performance is stable across the different market configuration considerered. \\\\
	\vspace{1cm}
	\begin{minipage}{\linewidth}
		\centering
		\captionof{table}{DDPG MSE Rewards} \label{tab:title3}
		\begin{tabular}{lrrrr}
			\toprule {} type&      Skew &  High Smile &  Inv. Smile \\ \midrule Bench          & -0.011913 &    -0.000022 &      -0.000044 \\
			seeds' avg & -0.017231 &     -0.005591 &    -0.001318    \\
			\bottomrule
		\end{tabular} 
	\end{minipage}
    \\\\
	\vspace{1cm}
	\begin{minipage}{\linewidth}
		\centering
		\captionof{table}{SAC MSE Rewards} \label{tab:title4}
		\begin{tabular}{lrrrr}
			\toprule {} type&      Skew &  High Smile &  Inv. Smile \\ \midrule Bench          & -0.011913 &    -0.0000220 &      -0.0000436 \\
			seeds' avg & -0.017717  &   -0.0006141   &       -0.0018047  \\
			\bottomrule
		\end{tabular} 
	\end{minipage}

	\subsubsection{Skew Market (MSE)}

	\begin{figure}[H]
		\centering
		\includegraphics[width=0.5\textwidth]{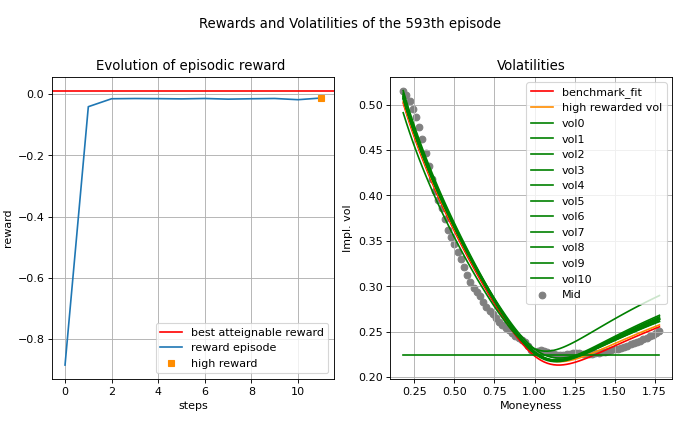}
		\caption{\label{fig:skew} Episodic agent's Rewards and Implied volatility evolution in a ''Skewed'' configuration with \textbf{DDPG}.}
	\end{figure}
	The figure shows that, the agent  is detecting the right parameter shifts from the first step. 
	\begin{figure}[H]
		\centering
		\includegraphics[width=0.5\textwidth]{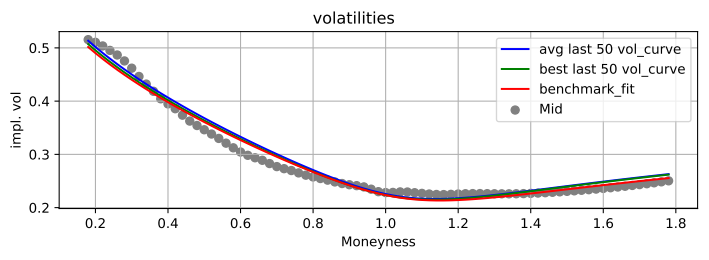}
		\caption{\label{fig:skew} Implied volatility snapshot in a ''Skewed'' configuration with \textbf{DDPG}: A Monte-Carlo on 5 differents random seeds is performed with a power decaying noise. We represent in ($ \textcolor{green}{green} $) the best response of the agent amongst the last 50 episodes and in ($ \textcolor{blue}{blue} $) the mean of the final volatility slices for the last 50 episodes.}
	\end{figure}

    \begin{figure}[H]
    	\centering
    	\includegraphics[width=0.5\textwidth]{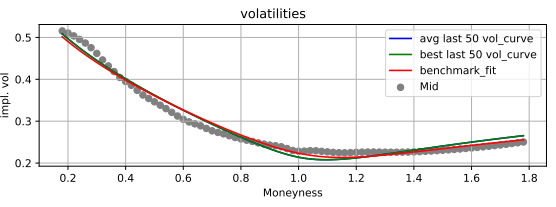}
    	\caption{\label{fig:skew} Implied volatility snapshot in a ''Skewed'' configuration with \textbf{SAC}: A Monte-Carlo on 5 differents random seeds is performed with automatic entropy adjustement. We represent in ($ \textcolor{green}{green} $) the best response of the agent amongst the last 50 episodes and in ($ \textcolor{blue}{blue} $) the mean of the final volatility slices for the last 50 episodes.  }
    \end{figure}

    \subsubsection{Smile Market (MSE)}

    \begin{figure}[H]
    	\centering
    	\includegraphics[width=0.485\textwidth]{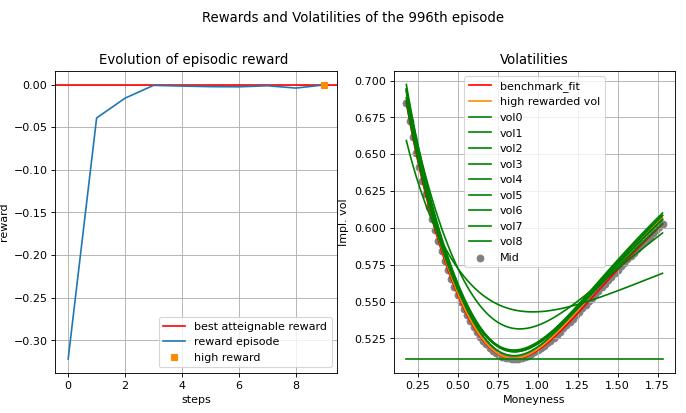}
    	\caption{\label{fig:smile} Episodic agent's Rewards and Implied volatility evolution in a ''Smile'' configuration with \textbf{DDPG}.}
    \end{figure}
    The figure shows that, the agent  is detecting the right parameter shifts from the first step.
    
    \begin{figure}[H]
    	\centering
    	\includegraphics[width=0.5\textwidth]{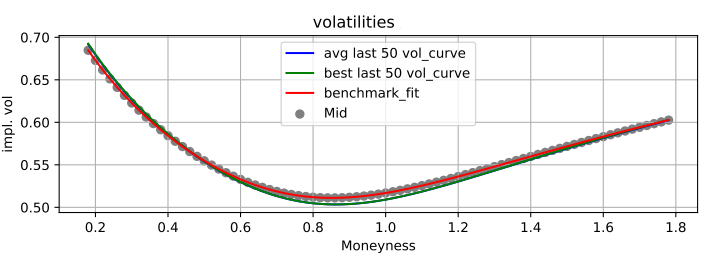}
    	\captionof{figure}{\label{fig:smile} Implied volatility snapshot in a ''High Smile'' configuration with \textbf{DDPG}: A Monte-Carlo on 5 different random seeds is performed with a power decaying noise. We represent in ($ \textcolor{green}{green} $) the best response of the agent amongst the last 50 episodes and in ($ \textcolor{blue}{blue} $) the mean of the final volatility slices for the last 50 episodes. 
    	}
    \end{figure}

    \begin{figure}[H]
    	\centering
    	\includegraphics[width=0.5\textwidth]{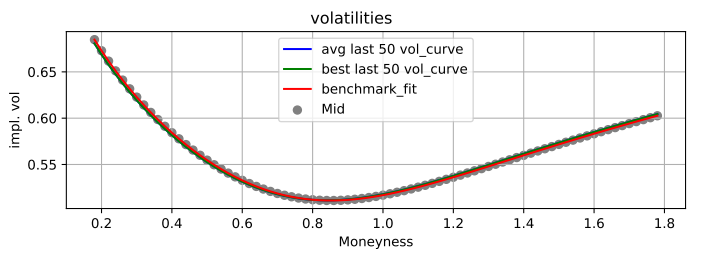}
    	\captionof{figure}{\label{fig:smile} Implied volatility snapshot in a ''High Smile'' configuration with \textbf{SAC}: A Monte-Carlo on 5 different random seeds is performed with automatic entropy adjustement. We represent in ($ \textcolor{green}{green} $) the best response of the agent amongst the last 50 episodes and in ($ \textcolor{blue}{blue} $) the mean of the final volatility slices for the last 50 episodes. 
    	}
    \end{figure}
   
   \subsubsection{Inverse Smile Market (MSE)}

   \begin{figure}[H]
   	\centering
   	\includegraphics[width=0.49\textwidth]{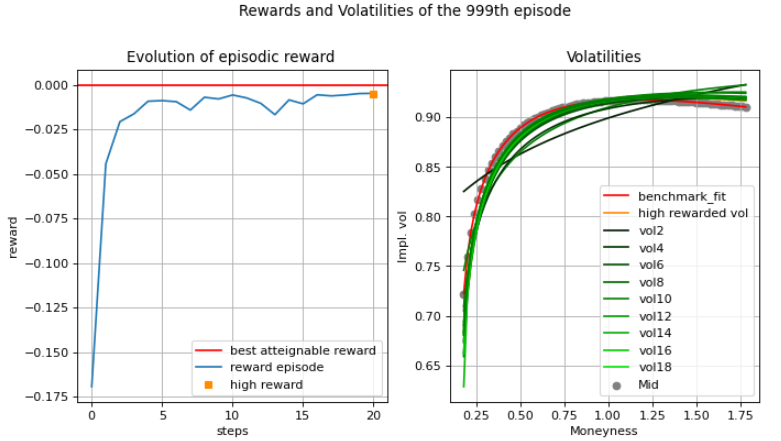}
   	\caption{\label{fig:smile} Episodic agent's Rewards and Implied volatility evolution in an ''Inverse Smile'' configuration with \textbf{DDPG}.}
   \end{figure}
   The figure shows that, the agent  is detecting the right parameter shifts from the first step.
   
   \begin{figure}[H]
   	\centering
   	\includegraphics[width=0.5\textwidth]{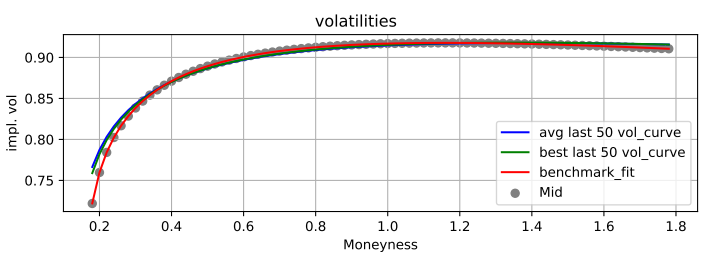}
   	\captionof{figure}{\label{fig:smile} Implied volatility snapshot in an ''Inverse Smile'' configuration with \textbf{DDPG}: A Monte-Carlo on 5 different random seeds is performed with a power decaying noise. We represent in ($ \textcolor{green}{green} $) the best response of the agent amongst the last 50 episodes and in ($ \textcolor{blue}{blue} $) the mean of the final volatility slices for the last 50 episodes. 
   	}
   \end{figure}
   
   \begin{figure}[H]
   	\centering
   	\includegraphics[width=0.5\textwidth]{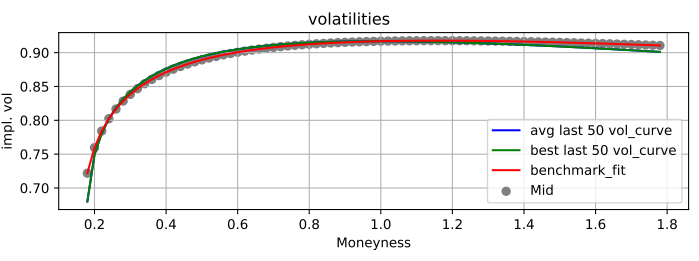}
   	\captionof{figure}{\label{fig:smile} Implied volatility snapshot in an ''Inverse Smile'' configuration with \textbf{SAC}: A Monte-Carlo on 5 different random seeds is performed with automatic entropy adjustement. We represent in ($ \textcolor{green}{green} $) the best response of the agent amongst the last 50 episodes and in ($ \textcolor{blue}{blue} $) the mean of the final volatility slices for the last 50 episodes. 
   	}
   \end{figure}

   For more details on the black-scholes vega weighted reward, please refer to the \textbf{appendix} \ref{appendix:subsection:appendix2}\\
   
   \vspace{.5cm}
  \textbf{Remarks:} SAC typically converges faster and more consistently across different environments. Moreover the introduction of entropy helps mitigate over fitting during training. While robust, its require handling both a stochastic policy and an entropy regularization term. In our settings where fine continuous control over actions is important, DDPG version might be a better choice since it may be more straightforward to integrate and tune for a quasi dynamic environment (which might be more sensitive to precise adjustments). It also has less computational overhead and resource usage.

   \subsection{Quasi-dynamic scenario}
   In this section we present the training, validation and testing phases of RL agents using \textbf{DDPG} algorithm. The agents are trained on data generated for two stocks (one with a wide spread and the other with a tight spread) across several episodes, each consisting of 50 steps. While the convergence rate of DDPG can be slow due to its sensibility to hyperparameters, Fine-tuning is often crucial for successful training.
    
\begin{figure}[H]
    		\centering
    		\includegraphics[width=0.45\textwidth]{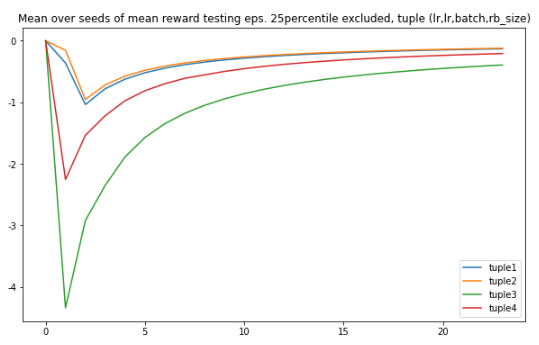}
    		\caption{\label{fig:Cumul_hyperparams_tuning} Cumulative mean rewards of evaluation episodes (for 5 tuples of hyper-parameters in the grid search).}
    	\end{figure}

   \paragraph{\underline{Training Phase:}}
        The agents are trained to maximize the average reward 
    	during evaluation episodes within a game. In those evaluation episodes, all sources of randomness in the DRL algorithms are turned off and no updates to the neural networks are made. We construct a hypercube of hyper-parameters (e.g: learning rates, volatility noise, replay buffer size, batch size, etc). For each tuple or hyper-parameter combination, we track the evolution of average rewards ($25\%$ quantile excluded) during the evaluation process. This method enables us to identify the optimal set of hyper-parameters, and we define the stopping criterion for training as the highest average reward obtained for the optimal configuration. (see \ref{fig:Cumul_hyperparams_tuning}).
    	
    	\begin{figure}[H]
    		\centering
    		\includegraphics[width=0.5\textwidth]{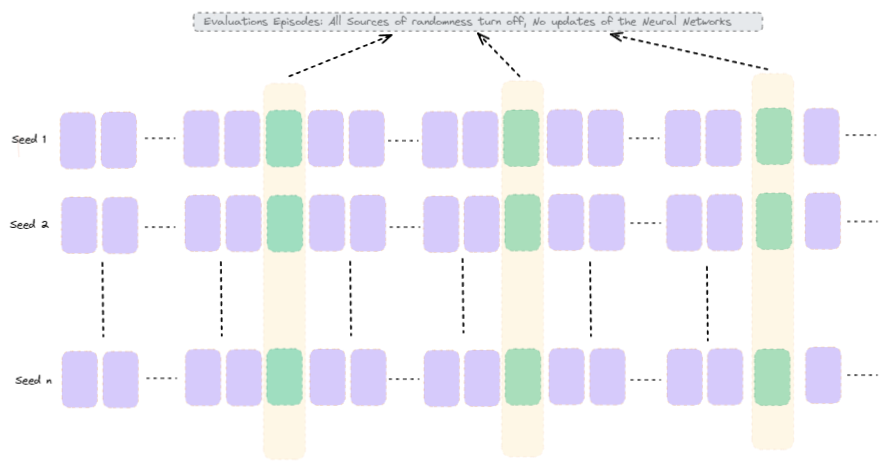}
    		\caption{\label{fig:hyperparams_tuning} Consistency of agent performance across multiple random seeds during the training phase.}
    	\end{figure}
    
    	\begin{figure}[H]
    		\centering
    		\includegraphics[width=0.45\textwidth]{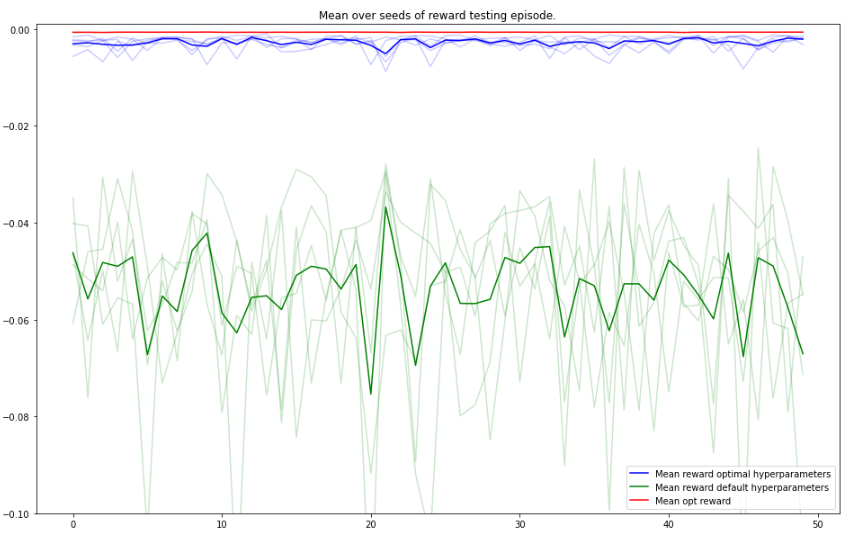}
    		\caption{\label{fig:hyperparams_tuning} Comparison of mean episodic rewards across multiple random seeds for agents trained with default versus optimized hyper-parameters.}
    	\end{figure} 
    
    As illustrated in Figure\ref{fig:hyperparams_tuning}, we compute the average evaluation rewards across multiple random seeds to identify the optimal set of hyper-parameters and establish the stopping criterion. The curves shown in Figure \ref{fig:Cumul_hyperparams_tuning} represent the mean results across seeds, ensuring a robust, stable and consistent process throughout the validation phase. 
    
    Once the optimal hyper-parameters are determined, we can compare the mean testing episodic rewards(averaged across a different set of random seeds) for agents trained with default versus optimized hyper-parameters.
    Figure \ref{fig:hyperparams_tuning} demonstrates the substantial improvement achieved when fine-tuning hyper-parameters compared to the default setup.
    
    \paragraph{\underline{Validation Phase:}}
         upon completion of training, we evaluate the performance of distinct RL agents using \textit{different random seeds} within a validation episode. 
   
    	\begin{figure}[H]
    		\centering
    		\includegraphics[width=0.49\textwidth]{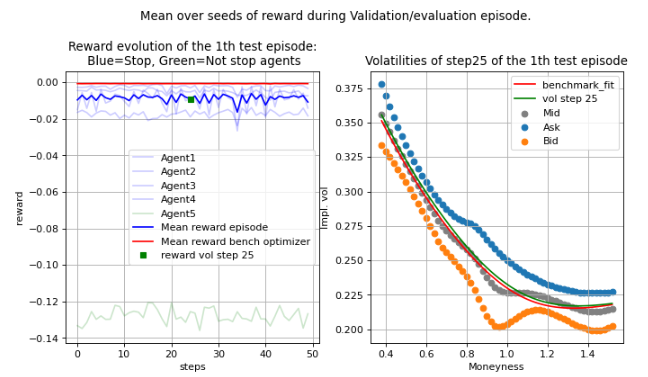}
    		\caption{\label{fig:wide_spread_validation} Evolution of mean episodic rewards(over several random seeds) and implied volatility(of one selected step) for a wide spread stock with \textbf{DDPG}. Notice that some agents (like Agent 5) don't reach the threshold. }
    	\end{figure}
    	
    	\begin{figure}[H]
    		\centering
    		\includegraphics[width=0.49\textwidth]{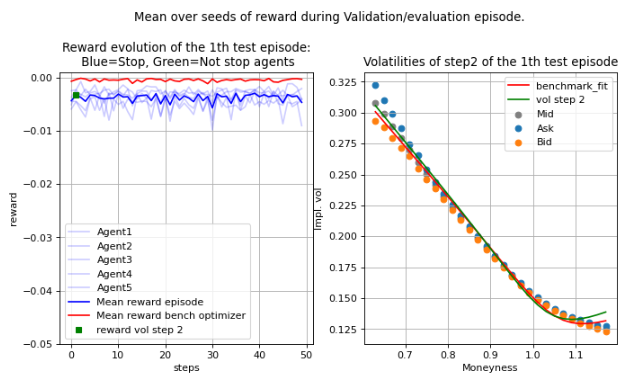}
    		\caption{\label{fig:tight_spread_validation} Evolution of mean episodic rewards(over several random seeds) and implied volatility(of one selected step) for a tight spread stock with \textbf{DDPG}.}
    	\end{figure}
\begin{figure}[H]
    		\centering
    		\includegraphics[width=0.52\textwidth]{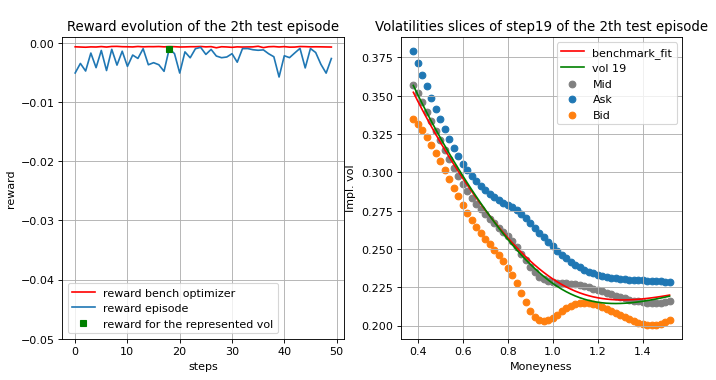}
    		\caption{\label{fig:wide_spread_testing} Evolution of agent's episodic rewards and implied volatility (of one selected step) for a \textbf{wide} spread stock during testing with \textit{DDPG}.}
    	\end{figure}
\begin{figure}[H]
    		\centering
    		\includegraphics[width=0.5\textwidth]{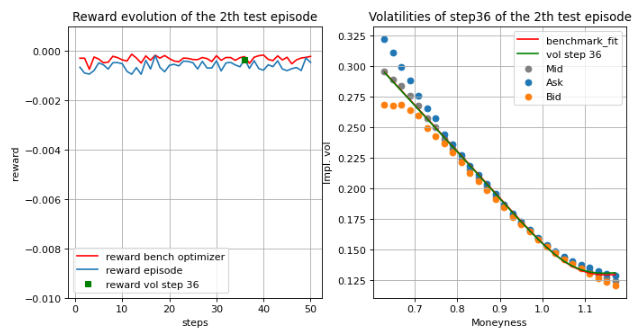}
    		\caption{\label{fig:tight_spread_testing} Evolution of agent's episodic rewards and implied volatility (of one selected step) for a \textbf{tight} spread stock during testing with \textit{DDPG}.}
    	\end{figure}

    We compute the mean reward for each agent, as shown in the figures below (\ref{fig:wide_spread_validation} and \ref{fig:tight_spread_validation}). 
    This validation phase allows us to identify agents that successfully reach or surpass the pre-determined reward threshold.
    \paragraph{\underline{Testing Phase:}}
    In the final testing phase, we select the agent with the best validation performance and assess its behavior in a test episode of 50 steps for different random seed. Figures \ref{fig:wide_spread_testing} and \ref{fig:tight_spread_testing} display the evolution of agent's rewards and implied volatility(of one selected step) during the testing phase for stocks with wide and tight spreads respectively.  
 	
 As can be seen from figures \ref{fig:wide_spread_testing} and \ref{fig:tight_spread_testing}, the performance of the best agent under both tigh and wide spread is satisfactory. The agent is able to track the market through time keeping the reward high (close to the optimal level for the number of parametrization coefficients used in the test). It is also important to highlight that there is also a good margin of improvement by deepening the hyperparametes search, extending the training period and optimization the scoring metric in the evaluation episodes. 

	\section{CONCLUSION AND PROSPECTS}\label{section:conclusion}
	In this paper, we proposed a model-free deep reinforcement learning architecture to solve the dynamic volatility fitting problem in continuous state and action spaces. We showed that DRL algorithms are natively	tailored for the volatility fitting problem as they possess (a) native exploration (b) effective catalog of past experiences and (c) good predictive power in large spaces. Using toy problems with increasing complexity,  	we showed that deep reinforcement learning algorithms can achieve  satisfactory performance.\\\\
	Traditional methods such as least squares fitting and gradient-based optimisation techniques, often struggle with complex state spaces and non-linearities. In contract RL's ability to learn from the interactions with the continuous environment allows it to progressively improve its performance by exploring and exploiting patterns that emerges from the data. Moreover, our approach highlights the benefits of rewards shaping and careful environment design, which allow the RL agent to learn efficiently in complex market conditions.\\\\
	This paper therefore lays the groundwork for a full AI-based dynamic volatility fitting. Future work can extend the fitting to larger parametrizations and reflects stylistic effects in volatility surface term-structure.

	\paragraph{ACKNOWLEDGMENTS:}

	The authors would like to thank all the members of Equities Derivatives Markets Quantitative Analysis (MQA) of Citigroup Global Market Limited for fruitful and insightful discussions, remarks and suggestions for this study. 
	
	We are particularly grateful to Truong Nguyen, Thomas Fouret and El Mostafa Ezzine for their feedback and encouragements throughout this 
	research project.

	\paragraph{DISCLAIMER:} 
	The article contains the personal views of the authors, which are not necessarily those of Citi. This is not a product of Citi Research.
	
	
	\newpage
	\bibliographystyle{siam}
	\bibliography{references}

	\appendix 

	\twocolumn
	
	\section{Black Scholes Vega Weighted MSE reward} \label{appendix:all}

	\subsection{Static case} \label{appendix:subsection:appendix1}
	As can be seen from table \ref{tab:title5} and \ref{tab:title6}, the final rewards achieved by the variants of the actor-critic algorithms is close to the one coming from the optimizer. The 
	performance is stable across the different market configuration considerered. \\\\
	\vspace{1cm}
	\begin{minipage}{\linewidth}
		\centering
		\captionof{table}{DDPG MSE Rewards} \label{tab:title5}
		\begin{tabular}{lrrrr}
			\toprule {} type&      Skew &  High Smile &  Inv. Smile \\ \midrule Bench          & -0.0217 &    -0.000041 &      -0.0000383 \\
			seeds' avg &  -0.0114 &     -0.001279 &       -0.0001314 \\
			\bottomrule
		\end{tabular} 
	\end{minipage}
	\\\\
	\vspace{1cm}
	\begin{minipage}{\linewidth}
		\centering
		\captionof{table}{SAC MSE Rewards} \label{tab:title6}
		\begin{tabular}{lrrrr}
			\toprule {} type&      Skew &  High Smile &  Inv. Smile \\ \midrule Bench          & -0.0217 &    -0.000041 &      -0.0000383 \\
			seeds' avg &  -0.0319 &     -0.003708 &     -0.000254    \\
			\bottomrule
		\end{tabular} 
	\end{minipage}

	\subsubsection{High Smile Market}
	\begin{figure}[H]
		\centering
		\includegraphics[width=0.45\textwidth]{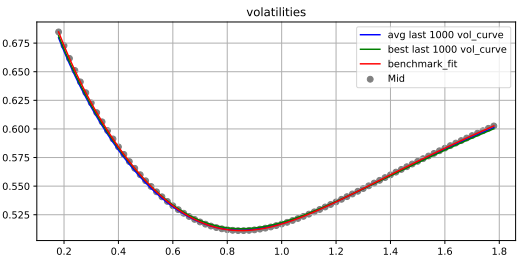}
		\captionof{figure}{\label{fig:smile} Implied volatility snapshot in a ''High Smile'' configuration with \textbf{DDPG}: A Monte-Carlo on 5 differents random seeds is performed with a power decaying noise. We represent in ($ \textcolor{green}{green} $) the best response of the agent amongst the last 1000 episodes and in ($ \textcolor{blue}{blue} $) the mean of the last 1000 volatility slices. 
		}
	\end{figure}

	\begin{figure}[!h]
		\centering
		\includegraphics[width=0.45\textwidth]{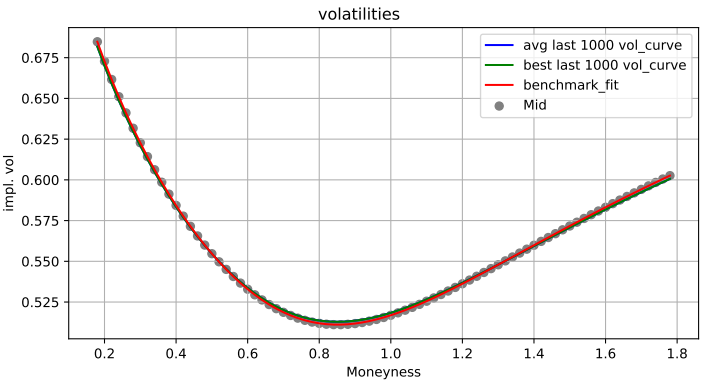}
		\captionof{figure}{\label{fig:smile} Implied volatility snapshot in a ''High Smile'' configuration with \textbf{SAC}: A Monte-Carlo on 5 differents random seeds. We represent in ($ \textcolor{green}{green} $) the best response of the agent amongst the last 1000 episodes and in ($ \textcolor{blue}{blue} $) the mean of the last 1000 volatility slices. 
		}
	\end{figure}

	\subsubsection{Skew Market}
	\begin{figure}[H]
		\centering
		\includegraphics[width=0.45\textwidth]{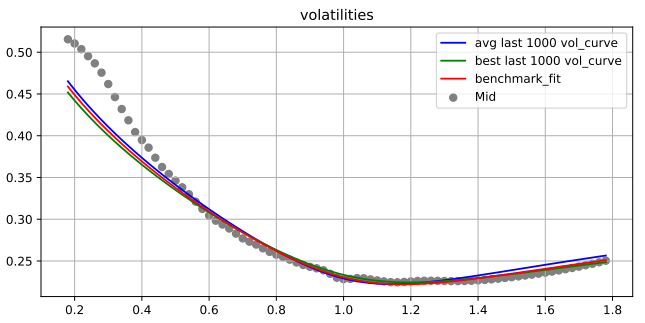}
		\caption{\label{fig:skew} Implied volatility snapshot in a ''skew Smile'' configuration with \textbf{DDPG}: A Monte-Carlo on 5 differents random seeds is performed with a power decaying noise. We represent in ($ \textcolor{green}{green} $) the best response of the agent amongst the last 1000 episodes and in ($ \textcolor{blue}{blue} $) the mean of the last 1000 volatility slices. }
	\end{figure}
	
	\begin{figure}[!h]
		\centering
		\includegraphics[width=0.45\textwidth]{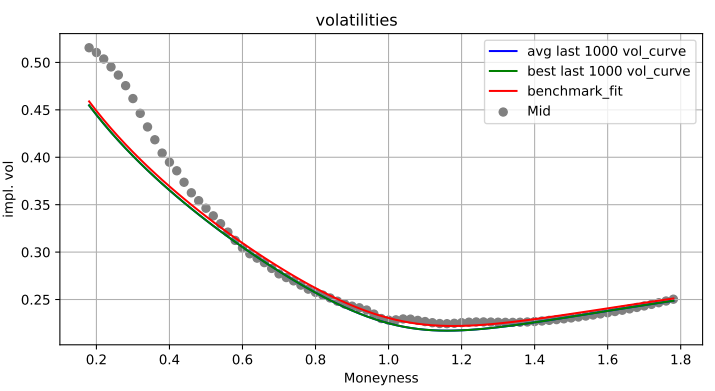}
		\captionof{figure}{\label{fig:smile} Implied volatility snapshot in a ''Skew'' configuration with \textbf{SAC}: A Monte-Carlo on 5 differents random seeds. We represent in ($ \textcolor{green}{green} $) the best response of the agent amongst the last 1000 episodes and in ($ \textcolor{blue}{blue} $) the mean of the last 1000 volatility slices. 
		}
	\end{figure}
	
	\subsubsection{Inverse Smile Market}
	
	\begin{figure}[H]
		\centering
		\includegraphics[width=0.41\textwidth]{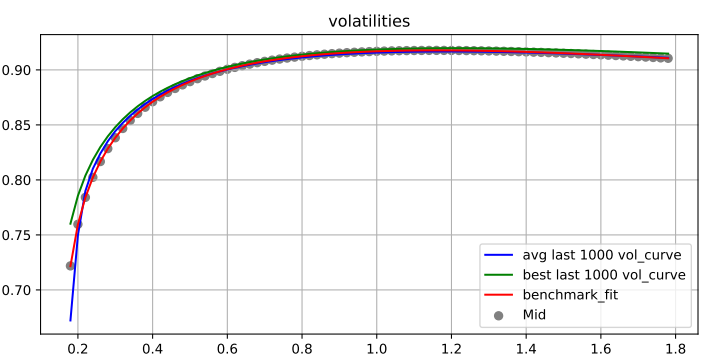}
		\caption{\label{fig:inversesmile} Implied volatility snapshot in an ''Inverse Smile'' configuration with \textbf{DDPG}: A Monte-Carlo on 5 differents random seeds is performed with a power decaying noise. We represent in ($ \textcolor{green}{green} $) the best response of the agent amongst the last 1000 episodes and in ($ \textcolor{blue}{blue} $) the mean of the last 1000 volatility slices. }
	\end{figure}

    \begin{figure}[!h]
    	\centering
    	\includegraphics[width=0.41\textwidth]{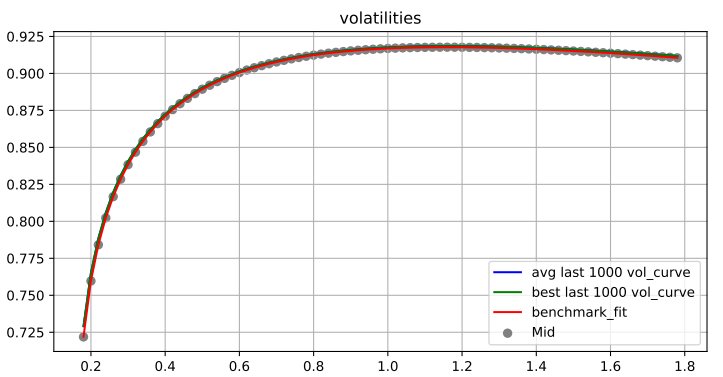}
    	\captionof{figure}{\label{fig:smile} Implied volatility snapshot in an ''Inverse Smile'' configuration with \textbf{SAC}: A Monte-Carlo on 5 differents random seeds. We represent in ($ \textcolor{green}{green} $) the best response of the agent amongst the last 1000 episodes and in ($ \textcolor{blue}{blue} $) the mean of the last 1000 volatility slices. 
    	}
    \end{figure}

	\subsection{Sequential case} \label{appendix:subsection:appendix2}
	
	As can be seen from table \ref{tab:title7}, the final rewards achieved by the variants of the actor-critic algorithms is close to the one coming from the optimizer. The 
	performance is stable across the different market configuration considerered. \\\\
	\vspace{1cm}
	\begin{minipage}{\linewidth}
		\centering
		\captionof{table}{DDPG MSE Rewards} \label{tab:title7}
		\begin{tabular}{lrrrr}
			\toprule {} type&      Skew &  High Smile &  Inv. Smile \\ \midrule Bench          & -0.0217 &    -0.000041 &      -0.0000383 \\
			seeds' avg &  -0.0791 &     -0.000818 &   -0.000516     \\
			\bottomrule
		\end{tabular} 
	\end{minipage}
	\\\\
	\vspace{1cm}

	\subsubsection{Smile Market}
	\begin{figure}[H]
		\centering
		\includegraphics[width=0.48\textwidth]{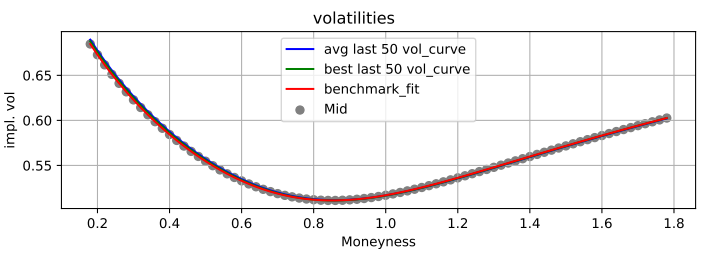}
		\captionof{figure}{\label{fig:smile} Implied volatility snapshot in a ''Smile'' configuration with \textbf{DDPG}: A Monte-Carlo on 5 differents random seeds is performed with a power decaying noise. We represent in ($ \textcolor{green}{green} $) the best response of the agent amongst the last 50 episodes and in ($ \textcolor{blue}{blue} $) the mean of the final volatility slices for the last 50 episodes. 
			}
	\end{figure}
   \subsubsection{Skew Market}
   \begin{figure}[H]
   	\centering
   	\includegraphics[width=0.48\textwidth]{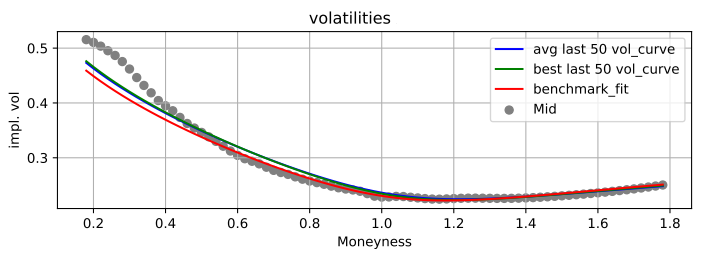}
   	\captionof{figure}{\label{fig:smile} Implied volatility snapshot in a ''Skew'' configuration with \textbf{DDPG}: A Monte-Carlo on 5 differents random seeds is performed with a power decaying noise. We represent in ($ \textcolor{green}{green} $) the best response of the agent amongst the last 50 episodes and in ($ \textcolor{blue}{blue} $) the mean of the final volatility slices for the last 50 episodes. 
   	}
   \end{figure}

      \subsubsection{Inverse Smile Market (MSE)}
   
   \begin{figure}[H]
   	\centering
   	\includegraphics[width=0.48\textwidth]{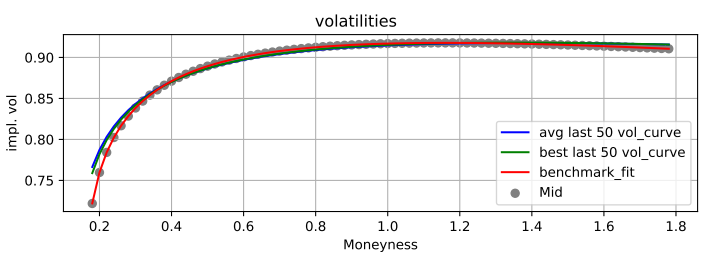}
   	\captionof{figure}{\label{fig:smile} Implied volatility snapshot in an ''Inverse Smile'' configuration with \textbf{DDPG}: A Monte-Carlo on 5 differents random seeds is performed with a power decaying noise. We represent in ($ \textcolor{green}{green} $) the best response of the agent amongst the last 50 episodes and in ($ \textcolor{blue}{blue} $) the mean of the final volatility slices for the last 50 episodes. 
   	}
   \end{figure}

    \newpage
    \section{Deep Reinforcement Learning}
    DRL comes from using neural networks thanks to the universal approximation theorem, as nonlinear functional approximation to the value and policy functions.
    
    \subsection{Fully Connected Artificial Neural Networks (FCNN)} \label{appendix:subsection:FCNN}
    
    FCNN refers to a neural network architecture where any given neuron is connected to all neurons in the previous layer. 
    
    Let us consider data samples $ x_i  \in  \mathbb{R}^{d_{in}}$ and $ y_i  \in  \mathbb{R}^{d_{out}}$  for $ 1 \leq i \leq M  $ associated to a regression problem, where  $ ( x_i) $ are the inputs and $ (y_i) $ are the outputs. That is, we look for a function $  F :\mathbb{R}^{d_{in}} \rightarrow \mathbb{R}^{d_{out}} $ which fits to the data i.e. $ \forall  1 \leq i \leq M , F(x_i) \sim y_i  $

    Let $ K + 1 $, $ K \in \mathbb{N} $ be the number of layers of the neural network and for $ k = 0, . . . ,K $; let $ d_k \in \mathbb{N} $ be the size of the $ k^{th} $ layer with $ d_0 = d_{in} $ and $ d_K = d_{out} $.
    
    Let $  \phi : \mathbb{R} \rightarrow \mathbb{R} $ be a non-linear function, generally chosen to be a sigmoid-type or a ReLU-type function (called activation function in the neural networks literature),  For $ k = 1, . . . , K, $ $ x  \in  \mathbb{R}^{d_{k-1}} $ and for $ \theta_k \in \mathcal{M}_{d_k,d_{k-1}} (( \mathbb{R}))$ (weight matrices) and $ b_k  \in \mathbb{R}^{d_k} $(bias vectors).  
    
    We define the below vector in $  \mathbb{R}^{d_k}  $ where the scalar function $ \phi $ is applied to the vector $  \theta_k · x + b_k $ coordinate by coordinate:
    
    \begin{align}
    \label{eq:acti}
    \phi_{\theta_k, b_k}(x) := [ \phi([\theta_k · x + b_k ]_i) ]_{1 \leq i \leq  d_k}
    \end{align}
    Writing $   \Phi  = (\theta_1, b_1, . . . , \theta_K, b_K) = (\theta_k, b_k)_{ k = 1, . . . , K} $, the output of the neural network with $ \sum_{l=1}^{K} d_{l-1} \times d_l $ parameters, is given by:
    
    \begin{align}
    \label{eq:fcnn}
    F_{\Phi}(x)  = \theta_K ·(\phi_{\theta_{K-1}, b_{K-1}} \circ  ...  \circ  \phi_{\theta_1, b_1}(u))+ b_K
    \end{align}
    
    \subsection{Training of Neural Networks} \label{appendix:subsection:FCNN_train}
     The objective is to extract a model from the empirical data. We look for a function $ F $ in a family of functions parametrized by a finite-dimension parameter: $ \left\lbrace F_{\Phi}, \Phi \in \mathbb{R}^{d_1 \times d_{in}} \times \mathbb{R}^{d_1}  \times...\times \mathbb{R}^{d_K \times d_{out}}  \times  \mathbb{R}^{d_K} \right\rbrace $
    
    For a loss function  $  L :\mathbb{R}^{d_{out}} \rightarrow \mathbb{R}^{+} $, which measures the error between the prediction $  F_{\Phi}(x_i)  $ and the true data $ y_i $, the regression problem is the minimization of the average loss over a small batch of data $ \mathcal{I}_{n+1} $. The parameters $ \Phi $ are updated in the descent direction of L using either Stochastic algorithms as introduced by Robbins and Monro \cite{Robbins1930Monro}   or a more stable version and still computationally efficient, the batch gradient descent and further their extension:
    
    \begin{align*}
    \label{eq:sgd}
    \Phi_{n+1}  = \Phi_{n} -  \frac{ \gamma_{n+1}}{N_{batch}}  \sum_{i  \in  \mathcal{I}_{n+1}} \partial_{\Phi}  F_{\Phi}(x_i) \dot \nabla L(F_{\Phi}(x_i) - y_i)
    \end{align*}
    
    Using Neural networks for reinforcement learning involve some (above-mentioned-like) optimization algorithm which assumes that samples are independently and identically distributed (IID). This assumption does not hold any more when the samples are generated from exploring sequentially in an environment, this is challenging in the context of RL. we used a Replay Buffer to address that issue and improve the learning performance. 
    
    \subsection{Replay Buffer} \label{appendix:subsection:ReplayMem}
    The agent learns over time to select his actions based on his past experiences (exploitation) and/or by trying new choices
    (exploration). The past experiences are stored in a \textit{''Replay Buffer''} mainly used to improve the performance of neural networks by providing nearly non correlated samples at each parameters update.
    Essentially, it is a finite size cache memory, which temporarily saves the agent observations during the learning process. 
    
    Before the training process, it is full of tuples or transitions $ (s_{t}, a_{t}, r_{t}, s_{t+1} ) $  sampled from the environment according to the initial exploration policy.
    During the training process and at each time step, we update the replay buffer by discarding the transitions with the worst reward or the oldest reward and storing a new tuple $ (s_{t}, a_{t}, r_{t}, s_{t+1} ) $ sampled from the environment according to the current policy. In this work, the weights of our Neural Networks are initialized from uniform distributions (Xavier initialisation).
    
    The first motivation of the replay memory in DQN \cite{mnih2015human} was to alleviate the problems of correlated data and non-stationary distributions by smoothing the training distribution over many past experiences. In fact, learning processes are usually in mini-batches i.e. the parameters of the neural networks are updated by uniformly choosing a minibatch of samples from the buffer and performing batch gradient descent (a more stable version and still computationally efficient of Stochastic gradient descent). Such algorithm assume each batch sample to be (IID) from a fixed distribution. However, exploration from the environment clearly generates correlated samples, subject to distributional shift. Thus we consider a \textit{''Replay Memory''} with large enough size, to store past experiences and uniformly sample a mini batch at each update of Neural Networks parameters in order to minimize the correlations between samples allowing the algorithms to learn across a set of uncorrelated transitions. 
    
	\onecolumn
	
	\section{Supplementary Results}
	
	The below figures display the evolution of some important quantities during the  training of the RL agents in several market type and MSE reward: A Monte-Carlo on 5 differents random seeds is performed with a power decaying noise for \textbf{DDPG} and automatic entropy adjustment for \textbf{SAC}. We represent in ($ \textcolor{green}{green} $) the best response of the agent amongst the last 1000 steps and in ($ \textcolor{blue}{blue} $) the mean of the final volatility slices for the last 1000 steps.\\ As we can see, the reward is converging and the replay buffer is improving over time. Moreover, the exploration parameters (std noise and temperature coefficient) are decreasing as the agent is converging to the optimal solution.

	\begin{figure}[H]
		\centering
		\includegraphics[width=0.85\textwidth]{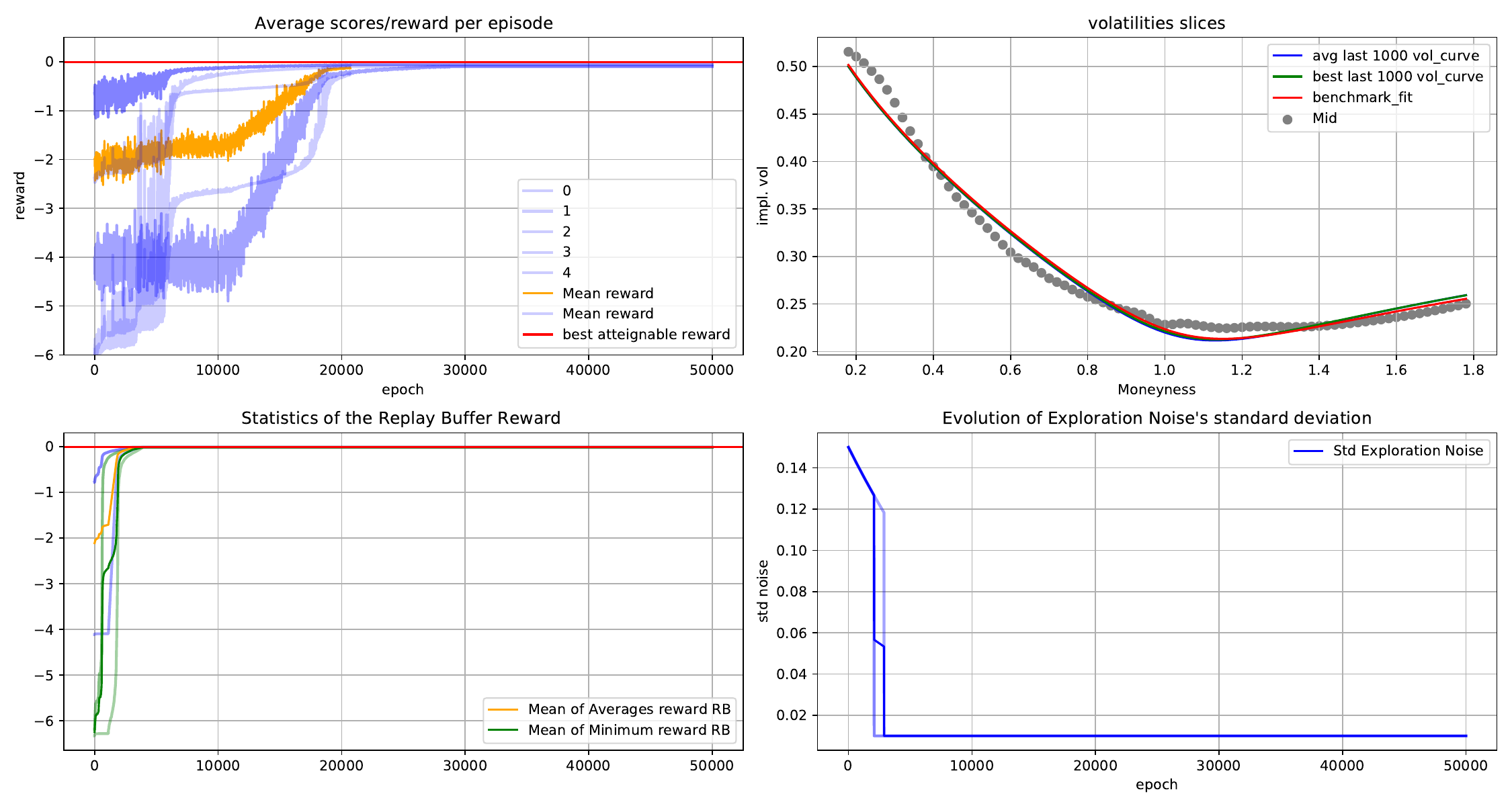} 
		\caption{\label{fig:smile_mkt} A snapshot of the training in a ''Skew'' configuration with \textbf{DDPG} and a power decaying noise.
		}
	\end{figure}
	\begin{figure}[H] 
		\centering
		\includegraphics[width=0.85\textwidth]{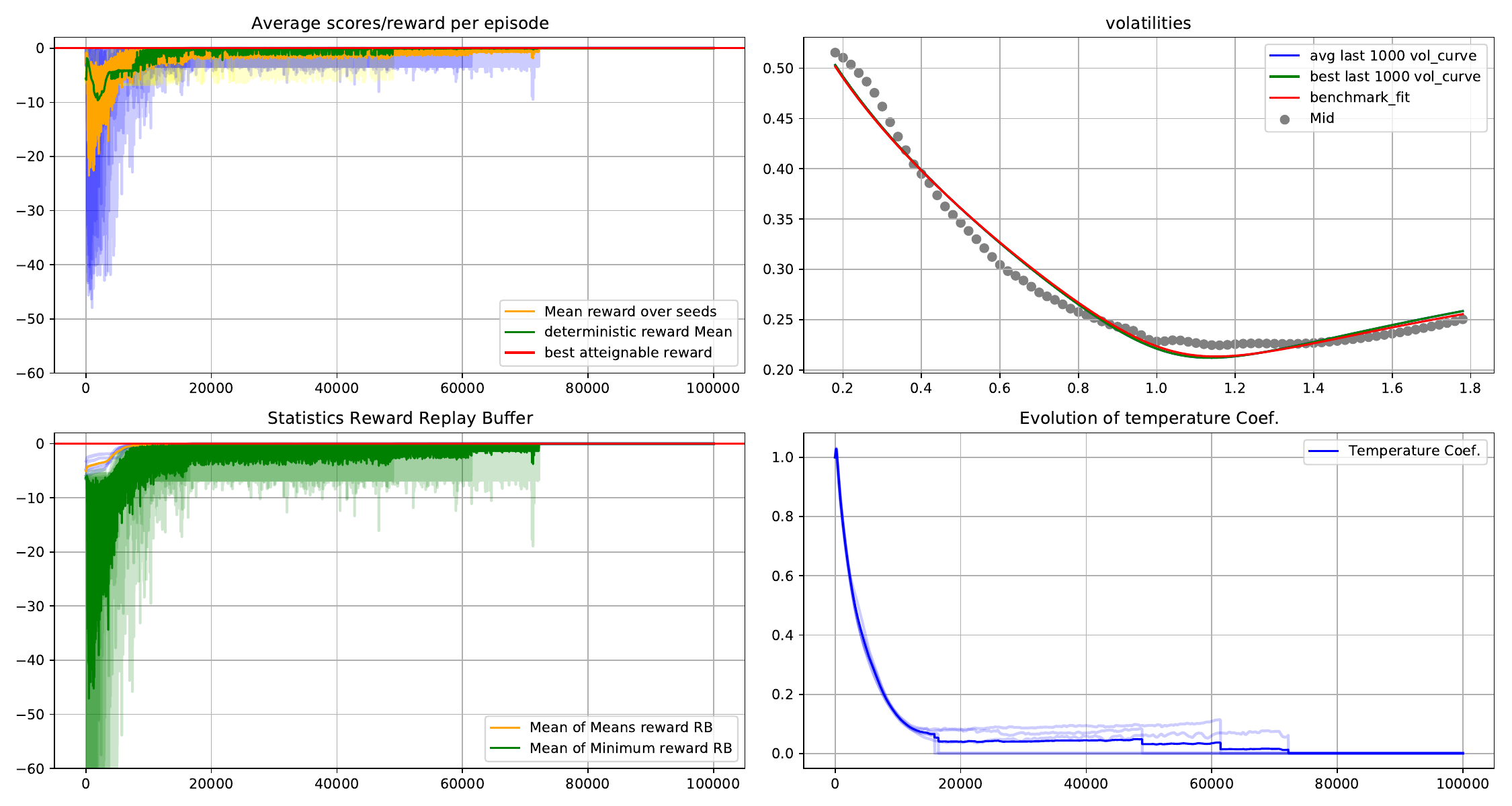}
		\caption{\label{fig:sac_highskew_mkt} A snapshot of the training in a ''Skew'' configuration with \textbf{SAC} and automatic entropy adjustement.
		}
	\end{figure}
	
	\begin{figure}[H]
		\centering
		\includegraphics[width=1\textwidth]{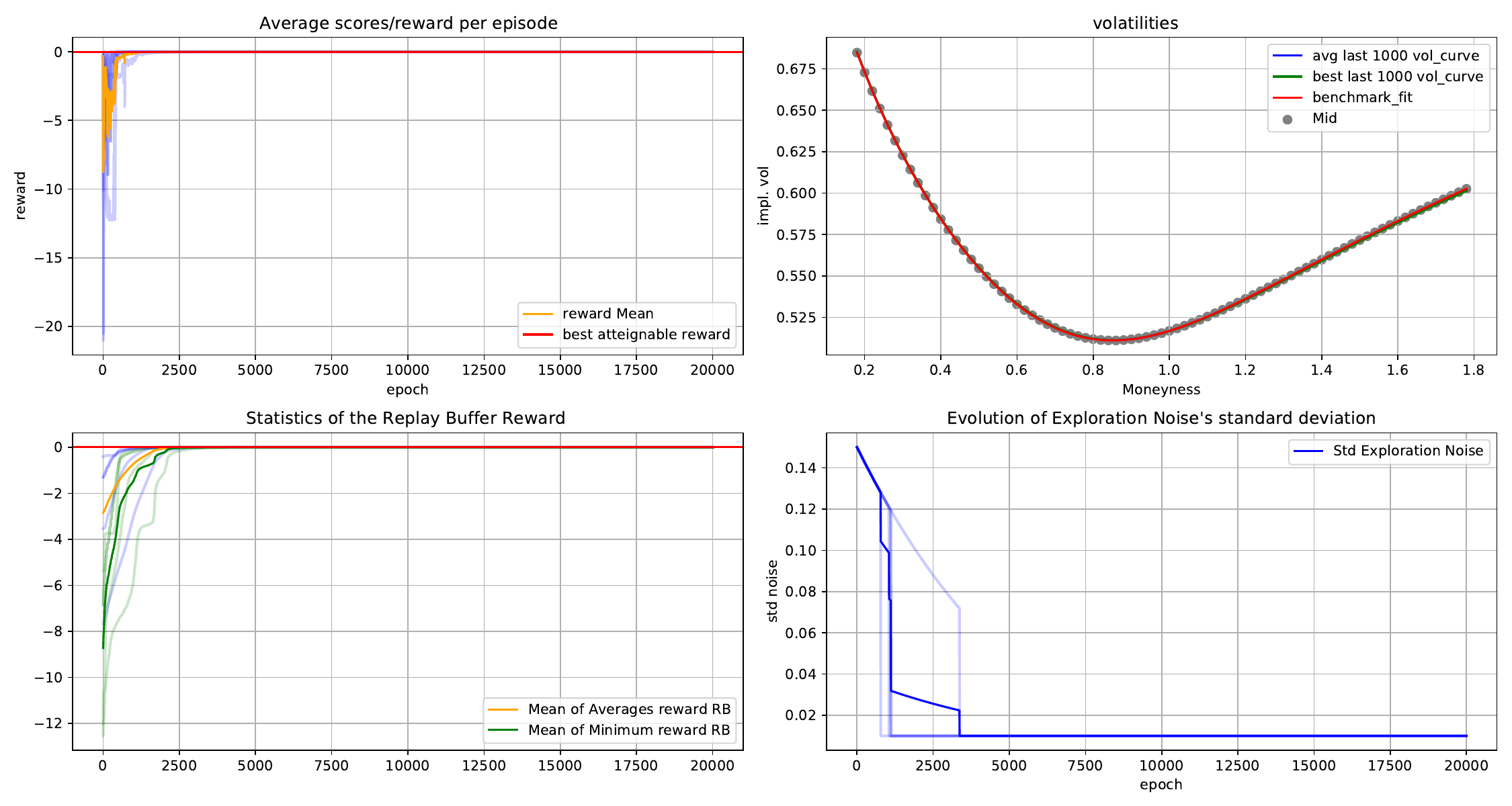}
		\caption{\label{fig:skew_mkt} A snapshot of the training in a ''High Smile'' configuration with \textbf{DDPG} and a power decaying noise. }
	\end{figure}
	
	\begin{figure}[H]
		\centering
		\includegraphics[width=1\textwidth]{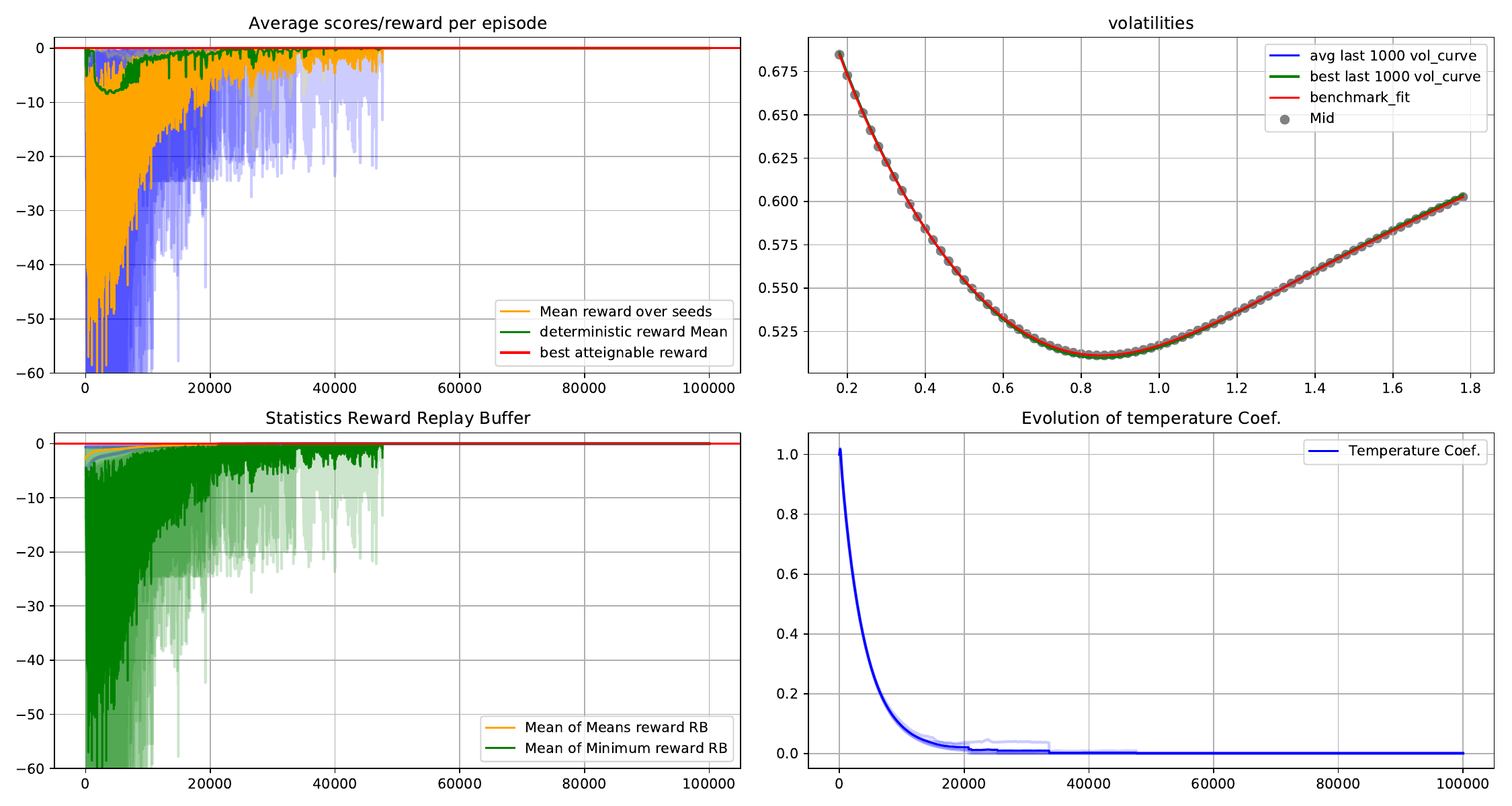}
		\caption{\label{fig:sac_smile_mkt}  A snapshot of the training in a ''High Smile'' configuration with \textbf{SAC} and automatic entropy adjustement.
		}
	\end{figure}

	\begin{figure}[H]
		\centering
		\includegraphics[width=1\textwidth]{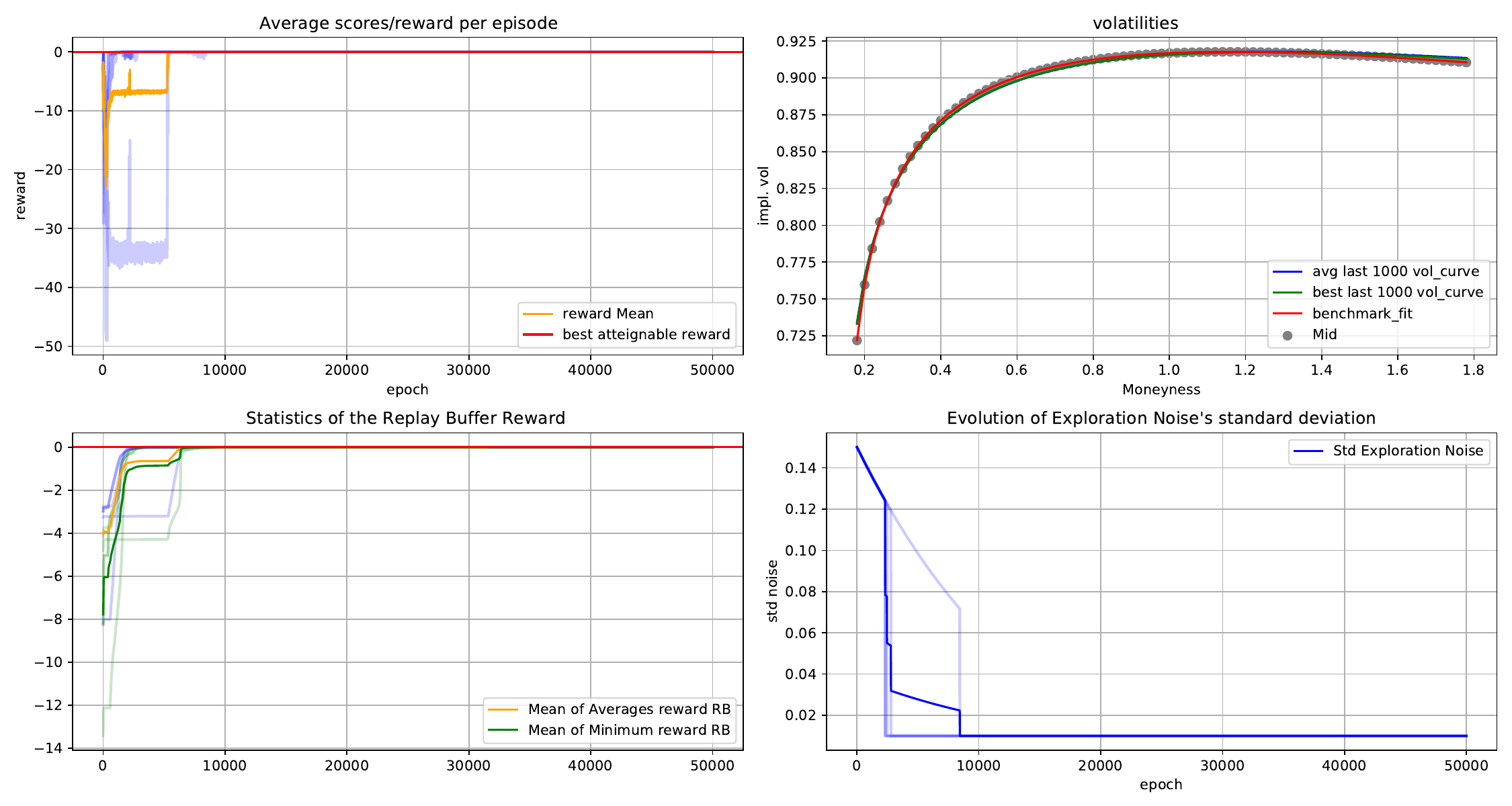}
		\caption{\label{fig:inversesmile_mkt}  A snapshot of the training in a ''Inverse Smile'' configuration with \textbf{DDPG} and a power decaying noise. }
	\end{figure}

    \begin{figure}[H]
    	\centering
    	\includegraphics[width=1\textwidth]{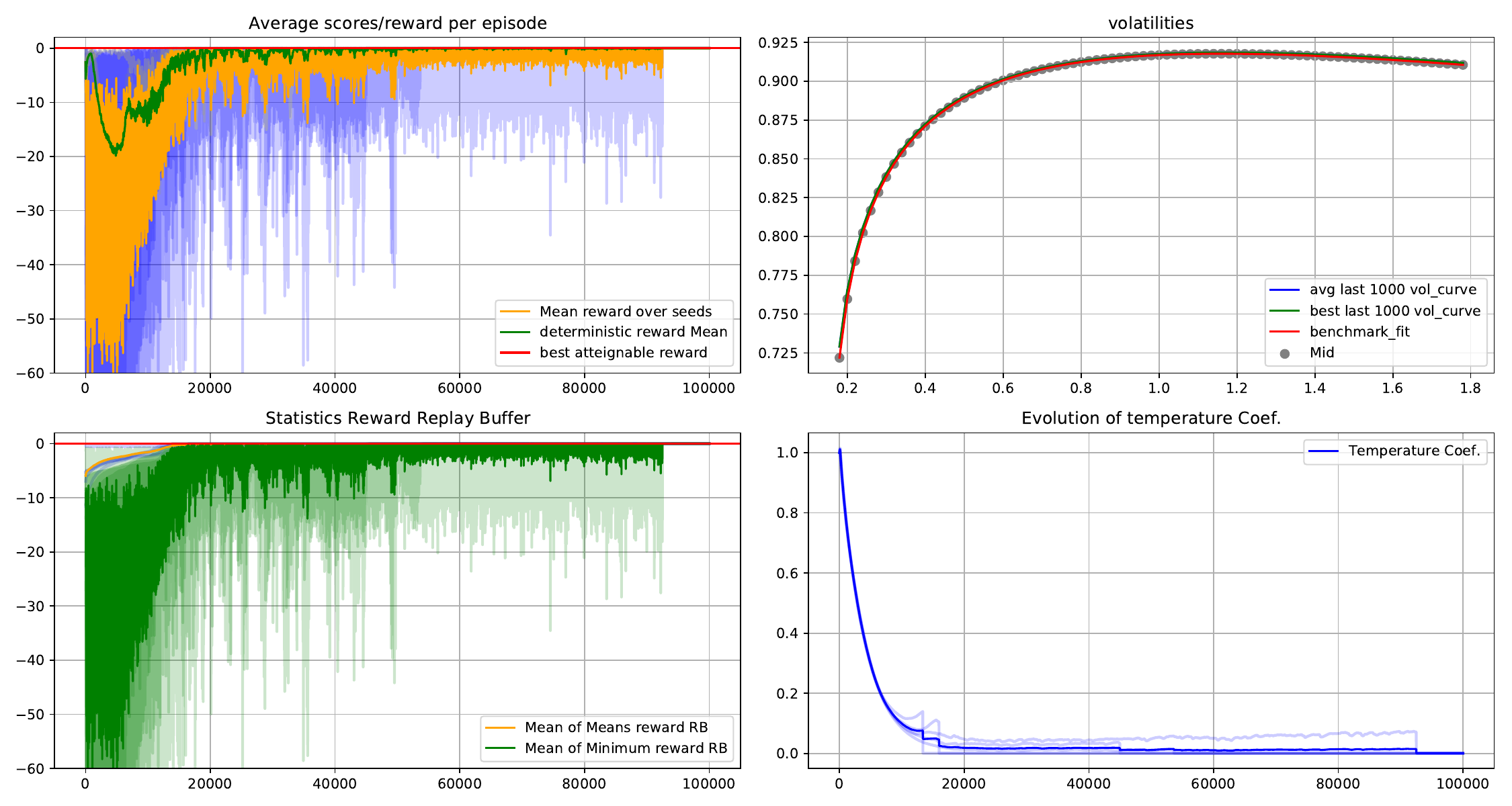}
    	\caption{\label{fig:sac_highskew_mkt} A snapshot of the training in a ''Inverse Smile'' configuration with \textbf{SAC} and automatic entropy adjustement.
    	}
    \end{figure}
 
 \newpage
 \section{Experiment Details and Hyperparameters} \label{appendix:section:Hyperparams}
 Below, we provide the different hyper-parameters during our experiments(Dynamic setting in brackets). 
 \subsection{DDPG variant Algorithm for Volatility fitting}
 \label{app:hypers1}
 \autoref{tab:params_ddpg} lists the DDPG parameters used in our experiments.
 \begin{table}[H]
 	\renewcommand{\arraystretch}{1.1}
 	\centering
 	\caption{DDPG Hyperparameters}
 	\label{tab:params_ddpg}
 	\vspace{1mm}
 	\begin{tabular}{l| l }
 		\toprule
 		Parameter &  Value\\
 		\midrule
 		optimizer &Adam \\
 		learning rate Actor and Critic & $0.0025$ ($ 2.5 \cdot 10^{-5}, 2.5 \cdot 10^{-4} $)\\
 		discount ($\gamma$) &  0.99\\
 		replay buffer size & $10^3$ ($2. 10^3$)\\
 		number of hidden layers (all networks) & 2\\
 		number of hidden units per layer & 256\\
 		number of samples per minibatch (batch size) & 64 (252)\\
 		Power decaying noise with std bounded in & $[0.01, 0.15]$ \\
 		nonlinearity &  ReLU, Tanh\\
 		target smoothing coefficient ($\tau$)& 0.001\\
 		gradient steps & 1\\
 		Neural Network initialisation & Xavier initialisation\\
 		\bottomrule
 	\end{tabular}
 \end{table}

 \subsection{SAC variant Algorithm for Volatility fitting}
 \label{app:hypers2}
 \autoref{tab:params_sac} lists the SAC parameters used in our experiments.
 \begin{table}[H]
 	\renewcommand{\arraystretch}{1.1}
 	\centering
 	\caption{SAC Hyperparameters}
 	\label{tab:params_sac}
 	\vspace{1mm}
 	\begin{tabular}{l| l }
 		\toprule
 		Parameter &  Value\\
 		\midrule
 		optimizer &Adam \\
 		learning rate Actor and Critic & $ 2.5 \cdot 10^{-5}, 2.5 \cdot 10^{-4} $\\
 		discount ($\gamma$) &  0.99\\
 		replay buffer size & $10^3$\\
 		number of hidden layers (all networks) & 2\\
 		number of hidden units per layer & 256\\
 		number of samples per minibatch (batch size) & 64\\
 		entropy target & $-\dim\left(\mathcal{A}\right) = - K$ \\
 		automatic entropy tuning & True \\
 		nonlinearity & ReLU, Tanh\\
 		target smoothing coefficient ($\tau$)& 0.001\\
 		Neural Network initialisation & Xavier initialisation\\
 		\bottomrule
 	\end{tabular}
 \end{table}
 \section{Hardware and Computational Ressources}
 All experiments in this paper were conducted on a machine with the following specifications: An Intel(R) Xeon(R) W-1270 CPU running at 3.40GHz with 64GB of RAM and 64-bit operating system in Windows 11.
 All algorithms were implemented in Python 3.8.8 using PyTorch 2.3.0+cpu. 
	
\end{document}